\documentclass[twocolumn, preprint2]{aastex631}

\usepackage{amsmath}

\pdfoutput=1

\received{June 19, 2023}

\revised{October 5, 2023}

\accepted{October 9, 2023}

\submitjournal{ApJ}

\shorttitle{JWST: D-, CN-PAHs and Overtone and Combination Bands}

\shortauthors{Boersma~et~al.}

\graphicspath{{./}{figures/}}

\begin{document}

\title{JWST: Deuterated PAHs, PAH-nitriles, and PAH Overtone and Combination
Bands \textit{I:\\ Program Description and First Look}}

\correspondingauthor{C.~Boersma}
\email{Christiaan.Boersma@nasa.gov}

\author[0000-0002-4836-217X]{C.~Boersma}
\affiliation{NASA Ames Research Center, MS 245-6, Moffett Field, CA 94035-1000, USA}

\author[0000-0002-6049-4079]{L.J.~Allamandola}
\affiliation{NASA Ames Research Center, MS 245-6, Moffett Field, CA 94035-1000, USA}

\author[0000-0001-6035-3869]{V.J.~Esposito}
\affiliation{NASA Ames Research Center, MS 245-6, Moffett Field, CA 94035-1000, USA}

\author[0000-0003-2552-3871]{A.~Maragkoudakis}
\affiliation{NASA Ames Research Center, MS 245-6, Moffett Field, CA 94035-1000, USA}

\author[0000-0002-1440-5362]{J.D.~Bregman}
\affiliation{NASA Ames Research Center, MS 245-6, Moffett Field, CA 94035-1000, USA}

\author[0000-0002-8341-342X]{P.~Temi}
\affiliation{NASA Ames Research Center, MS 245-6, Moffett Field, CA 94035-1000, USA}

\author[0000-0002-2598-2237]{T.J.~Lee}
\affiliation{NASA Ames Research Center, MS 245-6, Moffett Field, CA 94035-1000, USA}

\author[0000-0003-4716-8225]{R.C.~Fortenberry}
\affiliation{Department of Chemistry \& Biochemistry, University of Mississippi, MS 38677-1848, USA}

\author[0000-0002-2541-1602]{E.~Peeters}
\affiliation{Department of Physics and Astronomy \& The Institute for Earth and Space Exploration, The University of Western Ontario, London, ON N6A 3K7, Canada}
\affiliation{Carl Sagan Center, SETI Institute, 189 Bernardo Avenue, Suite 200, Mountain View, CA 94043, USA}

\begin{abstract}

A first look is taken at the NIRSpec 1-5~$\mu$m observations from \textit{JWST} program 1591 that targets 7 objects along the low-mass stellar life cycle with PAH emission. Spectra extracted from a 1.5$^{\prime\prime}$ radius sized circular aperture are explored, showing a wealth of features, including the 3~$\mu$m PAH complex, the PAH-continuum, and atomic and molecular emission lines from HI, He, H$_{\rm 2}$, and other species. CO$_{\rm 2}$- and H$_{\rm 2}$O-ice absorption and CO emission is also seen. Focusing on the bright-PDR position in M17, the PAH CH stretch falls at 3.29~$\mu$m (FWHM=0.04~$\mu$m). Signs of its 1.68~$\mu$m overtone are confused by line emission in all targets. Multi-component decomposition reveals a possible aliphatic deuterated PAH feature centered at 4.65~$\mu$m (FWHM=0.02~$\mu$m), giving [D/H]$_{\rm alip.}$=31$\pm$12.7\%. However, there is little sign of its aromatic counterpart between 4.36-4.43~$\mu$m. There is also little sign of PAH-nitrile emission between 4.34-4.39~$\mu$m. A PAH continuum rises from $\sim$1 to 3.2~$\mu$m, after which it jumps by about a factor of 2.5 at 3.6~$\mu$m, with bumps at 3.8, 4.04, and 4.34~$\mu$m adding structure. The CO$_{\rm 2}$ absorption band in M17 is matched with 10:1 H$_{\rm 2}$O:CO$_{\rm 2}$ ice at 10~K. The $v$=0 pure rotational molecular hydrogen population diagram reveals $>$2200~K UV-pumped gas. The hydrogen Pfund series runs from levels 10 to $>$30. Considering Br$\alpha$/Br$\beta$=0.381$\pm$0.01966 and Case B recombination results in A$_{\rm V}{\simeq}$8. CO emission in IRAS21282+5050 originates from 258~K gas. In-depth spectral-spatial analysis of all features and targets are planned for a series of forthcoming papers.

\end{abstract}


\section{Introduction}
\label{sec:introduction}

The James Webb Space Telescope \citep[\textit{JWST;}][]{2023PASP..135f8001G} launched Christmas morning 2021. After a roughly six month commissioning period \textit{JWST} started science operations. General Observer (GO) Cycle~1 program 1591 titled: ``\textit{NIRSpec IFU: Deuterated PAHs, PAH-nitriles, and PAH Overtone and Combination Bands}'' obtained its first set of observations September, 2022 and final set in May, 2023. This paper describes that program and takes a first, exploratory, look at the data.

The program set out to measure the 1-5~$\mu$m spectra of seven objects that represent key stages in the low-mass stellar life cycle and have strong polycyclic aromatic hydrocarbon (PAH) emission. The PAH model predicts many, but weak, features in this region that are able to resolve questions that have faced the PAH model since its inception. While ISO \citep[][]{1996A&A...315L..49D} and Akari \citep[][]{2007PASJ...59S.401O} have pioneered some of this spectral range, NIRSpec IFU's \citep[][]{2022A&A...661A..82B, 2023PASP..135c8001B} sensitivity and spectral resolution make it possible, for the first time, to measure the PAH signature with high fidelity across the entire region. 

Goals of the program are to quantify the amount of cosmological deuterium sequestered in PAHs; quantify the PAH deuterium/hydrogen-, and PAH carbon/nitrogen-ratios; track these ratios through the stellar life cycle; and provide insight into intermediate PAH structures, sizes, and charge states that determine how PAHs grow, erode, and evolve. The unique information contained in the 1-5~$\mu$m bands will forward our understanding of the interplay between PAHs, their environment and how they influence that same environment.

This paper is organized as follows. Section~\ref{sec:observations} starts out with describing the observations and detailing the data reduction. Section~\ref{sec:results} presents the results and Section~\ref{sec:discussion} provides a discussion. The paper finishes with a summary and conclusions in Section~\ref{sec:conclusions}.

\section{Observations}
\label{sec:observations}

\textit{JWST} GO Cycle~1 program 1591 targets seven well-studied astronomical PAH sources that cover key stages of the low-mass stellar life cycle. Within the confines of a small, 25-hour program, a careful balance was sought between obtaining a data set that would maximize discovery space, have a guaranteed science return, and not over-commit observing time. Table~\ref{tab:observations} summarizes the astrometric data of the targets, where it is noted that M17 was observed at two different positions and IRAS-21292+5050 has a separate, dedicated, background observation.

Utilizing the NIRSpec instrument in its IFU mode \citep[][]{2022A&A...661A..82B, 2023PASP..135c8001B}, a total of eight $\sim$1-5~$\mu$m medium resolution ($R\equiv\lambda/\Delta\lambda\simeq1000$) spatial-spectral observations were obtained, covering a field-of-view of $\sim$3.7$^{\prime\prime}\times$4.7$^{\prime\prime}$ at a pixel size of 0.1$^{\prime\prime}{\times}$0.1$^{\prime\prime}$. All observations use a 4-point dither and the G140M/F100LP (0.97-1.89~$\mu$m), G235M/F170LP (1.66-3.17~$\mu$m), and G395M/F290LP (2.87-5.27~$\mu$m) grating/filter combinations to cover 1-5~$\mu$m. Background (1 target) and LeakCal (3 targets) observations were limited to only a few targets and read-out-patterns were adjusted accordingly. Table~\ref{tab:observations} also summarizes the observational parameters used for each target.

\begin{deluxetable*}{lllllcccc}
  \tablecaption{Astrometric \& Observational Parameters\label{tab:observations}} \tablehead{
  \colhead{target} & \colhead{ra (ICRS)} &
  \colhead{dec (ICRS)} & \colhead{obs. date\tablenotemark{a}} &
  \colhead{type} & \colhead{groups\tablenotemark{b}} & \colhead{leakcal} & \colhead{pattern} & \colhead{exp.}}
  \startdata
      M17-PDR                        & 18 20 23.26 & -16 12 31.28 & Sep. 7,  2022 & H~\textsc{ii} &  4 & x & NRSIRS2      & 4.45h \\
      M17-B-PDR\tablenotemark{c}        & 18 20 24.35 & -16 11 52.96 & Sep. 9,  2022 & H~\textsc{ii} &  8 & x & NRSIRS2      & 6.22h \\
      NGC1333-SVS3                  & 03 29 10.40 & +31 21 51.12 & Sep. 17, 2022 & RN            & 10 & - & NRSIRS2RAPID & 2.05h \\
      NGC2023-PDR\tablenotemark{d}  & 05 41 41.00 & -02 15 41.70 & Oct. 12, 2022 & RN            & 20 & x & NRSIRS2RAPID & 4.15h \\
      NGC2023-PDR\tablenotemark{e}  & 05 41 41.00 & -02 15 41.70 & Feb. 19, 2023 & RN            & 20 & x & NRSIRS2RAPID & 4.15h \\
      IRAS21282+5050-RIM             & 21 29 58.69 & +21 29 58.69 & Sep. 16, 2022 & PPN           & 10 & - & NRSIRS2RAPID & 1.42h \\
      IRAS21282+5050-BKG             & 21 29 57.99 & +51 03 50.92 & Sep. 16, 2022 &               & 10 & - & NRSIRS2RAPID & 1.42h \\
      HD44179-SPIKE\tablenotemark{d} & 06 19 57.98 & -10 38 07.22 & Oct. 10, 2022 & PPN           &  9 & - & NRSIRS2RAPID & 1.81h \\
      HD44179-SPIKE\tablenotemark{e} & 06 19 57.98 & -10 38 07.22 & Feb. 19, 2023 & PPN           &  9 & - & NRSIRS2RAPID & 1.81h \\
      BD+303639-RIM                  & 19 34 45.28 & +30 30 52.72 & Sep. 20, 2022 & PN            &  9 & - & NRSIRS2RAPID & 1.99h \\
      NGC7027-EDGE                  & 21 07 01.50 & +42 14 20.79 & May. 25, 2023 & PN            & 10 & - & NRSIRS2RAPID & 1.86h \\
  \enddata
  \tablenotetext{a}{Data are proprietary in MAST for one year after being observed}
  \tablenotetext{b}{All observations were taken using a single integration}
  \tablenotetext{c}{Originally named M17-MC, but adjusted here, as the position clearly probes the bright PDR}
  \tablenotetext{d}{Observation affected by a short in the micro-shutter assembly (MSA)}
  \tablenotetext{e}{Re-take of observation affected by a short in the MSA per Webb Operations Problem Report (WOPR 88608)}
\end{deluxetable*}

\subsection{Targets}
\label{subsec:targets}

The seven object targeted are well-studied, have PAH emission, and probe different stages along the low-mass stellar life cycle. Each object is briefly discussed below, following the stages along the low-mass stellar life cycle, starting with H~\textsc{ii}-regions.\\

\paragraph{M17}\object{M17} is located in Sagittarius at a distance of 1800~pc \citep[][]{2005A&A...438.1163K}. It is one of the most massive known nearby star-forming regions \citep[][]{2009ApJ...696.1278P} and is irradiated by seven O stars, the most luminous being the binary system CEN1 with an effective temperature of 41200~K \citep[][]{1980A&A....91..186C}.

\begin{figure}
    \centering
    \includegraphics[width=\linewidth]{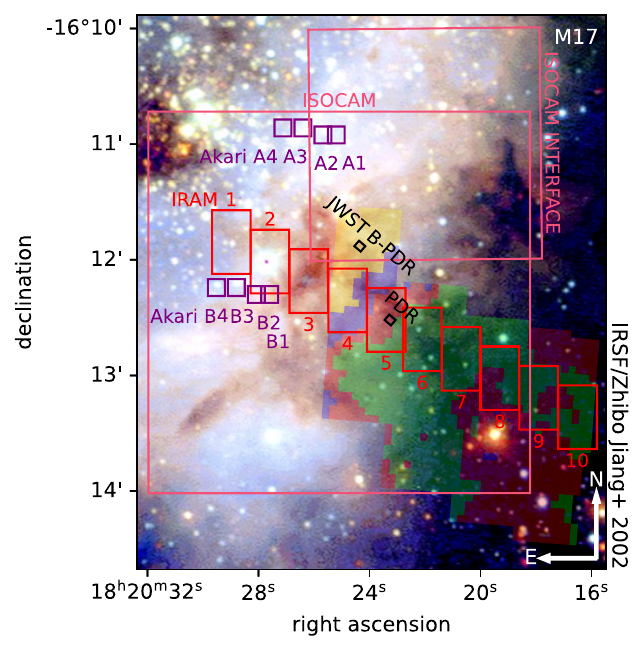}
    \caption{IRSF/SIRIUS three-color composite image of M17, with J-band data shown in blue, H-band data in green, and K$_{s}$-band data in red. Image credits: \citet{2002ApJ...577..245J}. Superimposed are telescope/instrument apertures from ISOCAM (pink), ISO-SWS (red), and Akari (purple) observations. The two \textit{JWST} apertures are shown and labeled in black. The translucent colored patches mark \textit{Spitzer}-IRS spectral map observations, where each patch represents one of four distinct emission zones identified through hierarchical clustering \citep[see][]{2018ApJ...858...67B}.}
    \label{fig:m17}
\end{figure}

Figure~\ref{fig:m17} shows a three-color composite image obtained by the SIRUS \citep[][]{1999sf99.proc..397N} instrument at the \textit{IRSF} 1.4-meter telescope. Overlain are a number of telescope apertures, including ten from ISO-SWS that trace the emission from the H~\textsc{ii}-region, crossing the PDR all the way into the molecular region \citep[][]{1996A&A...315L.337V}. Also shown are two ISOCAM \citep[][]{1999ESASP.427..663C} and eight Akari apertures \citep[][]{2016A&A...586A..65D} that focus on the H~\textsc{ii}-region. The footprint of \textit{Spitzer}-IRS spectral map observations are shown as four translucent colored patches, where each patch represents a distinct emission zone. These zones were identified through hierarchical clustering on the integrated intensity normalized spectra and reflect changes in the shape of the spectrum \citep[see][]{2018ApJ...858...67B}.

Two positions were targeted in M17. The first, labeled \textit{JWST}-PDR, is centered on the location that has a tentative detection of deuterated PAHs \citep[][]{2004ApJ...604..252P}. The second, labeled here \textit{JWST}-B-PDR, probes the bright PDR.

Dependent on the exact location and employed (emission) model, \citet{2004ApJ...604..252P} determine an aliphatic-to-aromatic ratio of 0.36$\pm$0.08 for this source, while \citet{2014ApJ...780..114O} find 0.023$\pm$0.004, and \citet{2016A&A...586A..65D} 0.09$\pm$0.05.

\begin{figure}
    \centering
    \includegraphics[width=\linewidth]{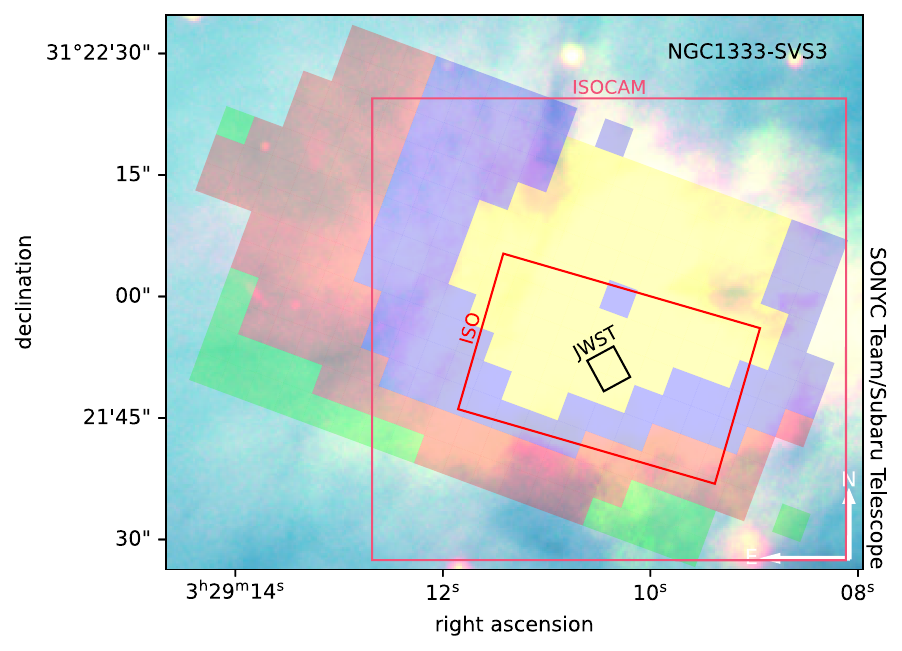}
    \caption{Three-color composite \textit{Subaru} image of part of NGC1333-SVS3. Here, the blue channel is taken from i$^{\prime}$-band, the green from J-band, and the red from K-band data. Image credits: SONYC Team/Subaru Telescope. Superimposed are apertures from ISOCAM (pink) and ISO-SWS (red) observations. The translucent colored patches mark \textit{Spitzer}-IRS spectral map data, where each color represents one of four zones identified through hierarchical clustering \citep[see][]{2016ApJ...832...51B}. The \textit{JWST} aperture is shown and labeled in black.}
    \label{fig:ngc1333-svs3}
\end{figure}

\paragraph{NGC1333-SVS3}\object{NGC1333} is located in Perseus, some 330~pc from Earth \citep{1970IAUS...38..219R}. SVS3 is the reflection nebula (RN) in NGC1333 that is illuminated by a likely binary composed of a 15700~K and a 6810~K F2 star \citep[][]{2002BaltA..11..261S}. Figure~\ref{fig:ngc1333-svs3} presents the morphology of the SVS3 region as seen by the 8.2 meter \textit{Subaru} telescope. Overlain are a number of telescope apertures, including those from ISOCAM and ISO-SWS observations. Also overlain, marking \textit{Spitzer}-IRS spectral map observations, are translucent patches indicating four distinct emission zones identified through hierarchical clustering (see \citealt{2016ApJ...832...51B} and the description of M17). It is noted that the available \textit{Spitzer} spectral map data cover a substantially larger region, but were trimmed to only encompass that with strong PAH emission. The targeted \textit{JWST} position borders a nebulous outflow arc that can be seen in ISOCAM imagery.

NGC1333-SVS3 has been the subject of numerous studies \citep[e.g.,][and references therein]{1996ApJ...460L.119J, 1999ApJ...513L..65S, 2016ApJ...819...65S}. For a review see \cite{2008hsf1.book..346W}.

\begin{figure}
    \centering
    \includegraphics[width=\linewidth]{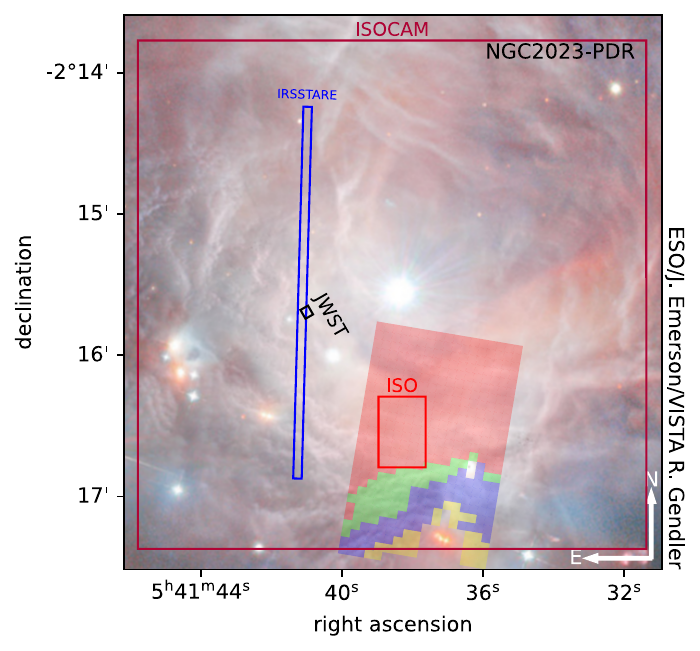}
    \caption{\textit{VISTA}-image of part of NGC2023. Image credits: ESO/J. Emerson/VISTA R. Gendler. Superimposed are telescope/instrument apertures for ISOCAM (pink), ISO-SWS (red), and \textit{Spitzer} (blue slit) observations. The colored translucent overlay marks \textit{Spitzer}-IRS spectral map observations, where each color represents one of four distinct emission zones identified through hierarchical clustering \citep[see][]{2016ApJ...832...51B}. The \textit{JWST} aperture is shown and labeled in black.}
    \label{fig:ngc2023}
\end{figure}

\paragraph{NGC2023}\object{NGC2023} is the RN located some 415~pc from Earth in Orion \citep[][]{2007A&A...474..515M}. It is hosting a cluster of faint, most likely, pre-main sequence stars, where the radiation environment is dominated by HD~37903, a B1.5 24800~K star \citep[]{2009A&A...507.1485M}. Figure~\ref{fig:ngc2023} presents the morphology of the region as seen by \textit{VISTA} \cite[][]{2006Msngr.126...41E, 2006SPIE.6269E..0XD}. Overlain are a number of telescope apertures, including those from ISOCAM \citep[][]{2002A&A...389..239A}, ISO-SWS \citep[][]{1999ESASP.427..727M}, and \textit{Spitzer} \citep[][]{2015ApJ...807...99A} single-slit observations. Also overlain, colored translucent patches marking \textit{Spitzer}-IRS spectral map observations depicting four emission zones identified through hierarchical clustering (see \citealt{2016ApJ...832...51B} and the description of M17). The \textit{JWST} position is the brightest spot in the nebula \citep[][]{2012ApJ...747...44P}.

PAH emission in NGC2023 has been the subject of numerous studies, including \cite{2012A&A...542A..69P}, \cite{2016ApJ...819...65S}, \citet{2016ApJ...832...51B}, \citet{2017ApJ...836..198P}, \citet{2019ApJ...887...46B}, and \citet{2021MNRAS.500..177S}.

\begin{figure}
    \centering
    \includegraphics[width=\linewidth]{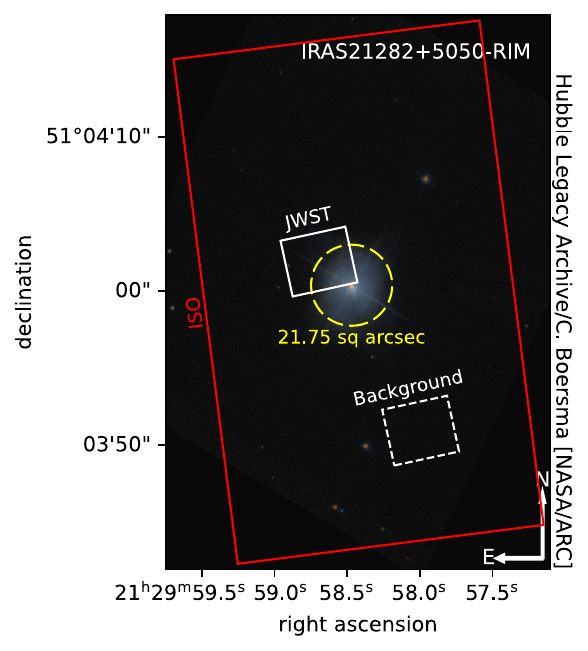}
    \caption{Hubble-ACS three-color composite image of IRAS21282+5050, with F814W data shown in red, F606W data in blue, and an equal mixture of F814W and F606W data in green. Superimposed is the aperture from ISO-SWS (red) observations and the apparent size of the nebula is indicated with the circle (yellow dashed). The \textit{JWST} apertures for the science target and separate, dedicated, background are shown and labeled in white, with the latter dashed.}
    \label{fig:iras21282}
\end{figure}

\paragraph{IRAS~21282+5050}\object{IRAS~21282+5050} is a young, $\sim$3000~yr old, very compact, low-excitation proto-planetary nebula \citep[PPN; ][]{1987ApJ...321L.151C, 1992ApJ...396..120B}. It is located 4000~pc from Earth \citep[][]{2020yCat.1350....0G} and hosts both a 28000~K post-AGB and a 6000~K main sequence star. Figure~\ref{fig:iras21282} shows a three-color composite Hubble image constructed from data obtained from the Hubble Legacy Archive overlain with the aperture of ISO-SWS observations \citep[][]{1996A&A...315L.369B}.

An aliphatic-to-aromatic band strength ratio of $\sim$0.47 has been determined for this source, suggesting early-stage photo-chemical processing \citep[][]{2019Ap&SS.364...32H, 2020ApJ...902..118S}. In addition, the $\sim$1.67~$\mu$m overtone of the 3.3~$\mu$m PAH band has been tentatively detected here \citep[][]{1993ApJ...408..586K, 1994ApJ...434L..15G, 2019A&A...632A..71C}.

The \textit{JWST} position borders the outskirts, avoiding the brightest parts of the nebula.

\paragraph{HD44179 (The Red Rectangle)}\object{HD44179} is a young, low-excitation PPN with a 23000~K mass-accretion driven radiation field hosting a 7700~K post-AGB and a $<$6000~K main sequence star \citep[][]{2009ApJ...693.1946W}. HD44179 has a complex geometry with a x-shaped outflow with density enhancements resembling rungs on a ladder.

\begin{figure}
    \centering
    \includegraphics[width=\linewidth]{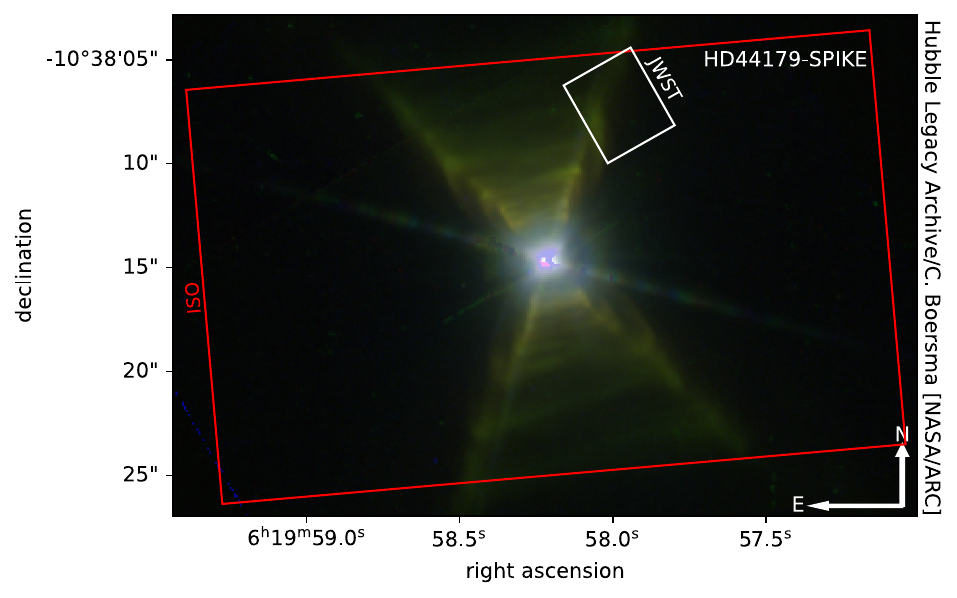}
    \caption{Three-color composite \textit{HST} WFPC2 image of HD44179, with the red channel taken from narrow band F631N data tracing [OI], the green channel from narrow band F588N data tracing HeI and the blue channel from medium Str\"{o}mgren b band F467M data. Superimposed is the aperture from ISO-SWS (red) observations. The \textit{JWST} aperture is shown and labeled in white.}
    \label{fig:hd44179}
\end{figure}

Figure~\ref{fig:hd44179} shows a three-color composite Hubble image constructed from data obtained from the Hubble Legacy Archive, overlain with the aperture of ISO-SWS observations \citep[][]{1998Natur.391..868W}. The \textit{JWST} aperture is located on one of the ``beams'' of the x-shaped ladder, where the ISOCAM spectrum is dominated by PAH emission.

The 3.3~$\mu$m PAH emission feature in HD44179 has been of particular interest, where it has been suggested to consist of two distinct spectroscopic components \citep[][]{2003MNRAS.346L...1S, 2007MNRAS.380..979S, 2012MNRAS.426..389C}. However, recently this has been disputed \citep[][]{2022ApJ...939...86T}.

\paragraph{BD+303639}\object{BD+303639} is a very young, $\sim$600~yr old, moderate excitation ($T_{\rm star}$$\simeq$47000~K) planetary nebula \citep[PN; ][]{1994MNRAS.271..449S}, located 1600~pc from Earth \citep[][]{2020yCat.1350....0G}. An aliphatic/aromatic ratio of $\sim$0.24 has been measured \citep[][]{2020ApJ...902..118S}. The faint structure near 3.4~$\mu$m in its ISO-SWS spectrum suggests PAH sculpting has started, where the least thermodynamically favored side groups are being removed \citep[][]{1996A&A...312..167L, 2020ApJ...902..118S}.

\begin{figure}
    \centering
    \includegraphics[width=\linewidth]{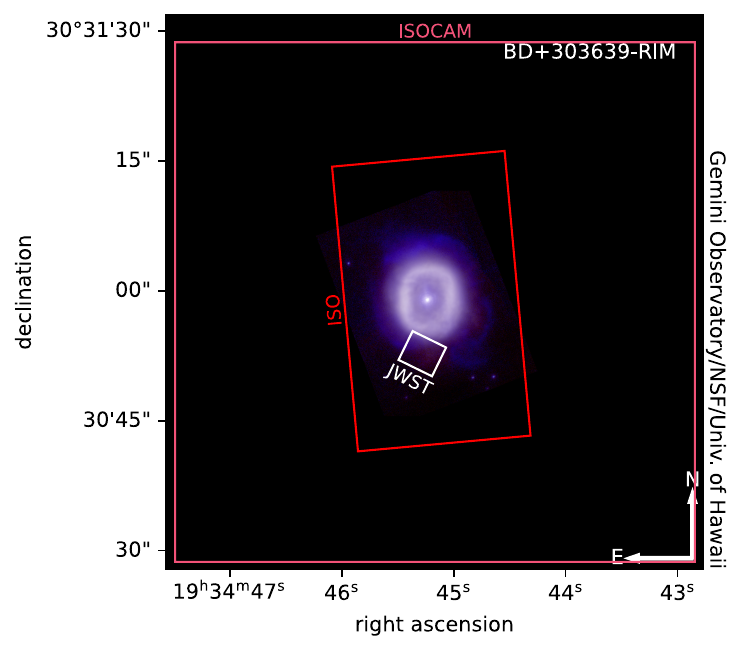}
    \caption{Three-color composite infrared adaptive optics image of BD+303639 from the Gemini North 8-meter telescope. The blue and red channels show narrow band H$_{\rm 2}$ and Br$\gamma$ emission, respectively. Image credits: Gemini Observatory, US National Science Foundation, and University of Hawaii Institute for Astronomy. Superimposed are apertures from ISOCAM (pink) and ISO-SWS (red) observations. The \textit{JWST} aperture is shown and labeled in white.}
    \label{fig:bd+303639}
\end{figure}

Figure~\ref{fig:bd+303639} shows the Gemini North adaptive optics image of the nebula. Overlain are telescope apertures from ISOCAM \citep[][]{1999A&A...351..201P} and ISO-SWS \citep[][]{2003A&A...406..165B} observations. The targeted position is located on the ring, where the ISOCAM spectrum shows strong PAH emission, but avoids the brightest parts of the nebula.

\begin{figure}
    \centering
    \includegraphics[width=\linewidth]{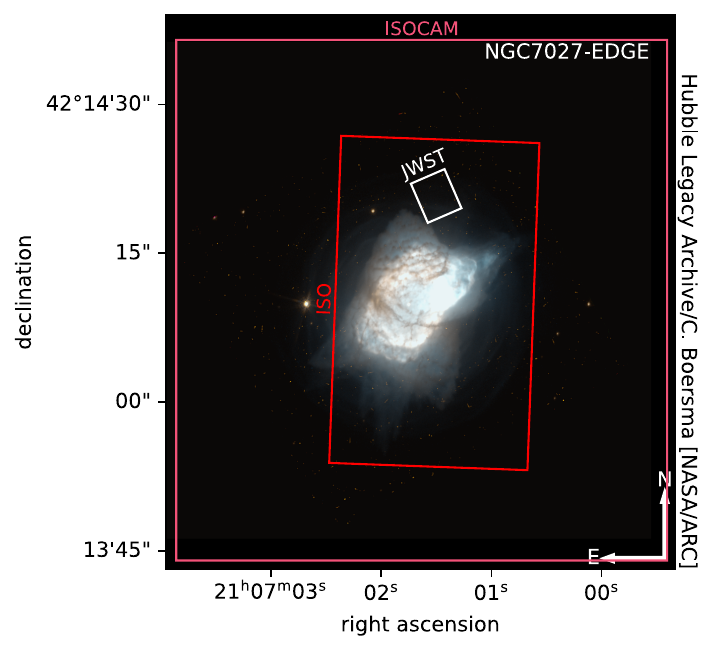}
    \caption{Three-color composite \textit{HST} image of NGC7027 with WFPC2 F55W data in red, F814W data in blue, and green a combination of the two. Superimposed are apertures of ISOCAM (pink) and ISO-SWS (red) observations. The \textit{JWST} aperture is shown and labeled in white.}
    \label{fig:ngc7027}
\end{figure}

\paragraph{NGC7027 (Jewel Bug Nebula)}\object{NGC7027} is a young, $\sim$700~yr, very high-excitation ($T_{\rm star}$$\simeq$198000~K) PN \citep[][]{1998Ap&SS.255..489C, 2000ApJ...539..783L}, located in Cygnus, some 920~pc from Earth. Figure~\ref{fig:ngc7027} shows the three-color composite Hubble image constructed from data obtained from the Hubble Legacy Archive, overlain with apertures from ISOCAM \citep[][]{1999A&A...351..201P} and ISO-SWS \citep[][]{1996A&A...315L.369B} observations. The targeted \textit{JWST} position lies on the outskirts of the nebula, where the ISOCAM spectrum shows a relatively moderate amount of PAH emission and the brightest regions of the nebula are avoided.

\subsection{Data Reduction}
\label{subsec:reduction}

Level-1b\footnote{STScI processed data: \dataset[https://doi.org/10.17909/x16g-g718]{https://doi.org/10.17909/x16g-g718}.} data associated with \textit{JWST} program 1591 were downloaded from MAST using the Python \texttt{astroquery} package. The \textit{JWST} \texttt{Detector1Pipeline} was run on the downloaded uncalibrated data using default parameters. Subsequently, \texttt{association} files were generated, ensuring proper tagging of LeakCal and Background observations. The \texttt{Spec2Pipeline} was run using default parameters, but skipping the \texttt{CubeBuild} and \texttt{Extract1D} steps. The \texttt{Spec2Pipeline} was run on the background exposures\footnote{The background of IRAS21282+5050-RIM caused the pipeline to issue a warning and skipping the background subtraction step.} treating them as science data. Next, \texttt{association} files were generated for combining the four dither positions. The \texttt{Spec3Pipeline} was then called with default parameters, but skipping the \texttt{OutlierDetection} and \texttt{Extract1D} steps. The Level-3 spectral cubes were build in the \texttt{ifualign} coordinate system, using Shepard's method with exponential weighting (\texttt{emsm}) for mapping the data into a rectilinear cube. For IRAS21282+5050-RIM, first the background exposures were subtracted from the level-2 calibrated science exposures. The JWST Calibration Pipeline at version 1.10.0 was used with CRDS context \texttt{jwst\_1077.pmap} running under Python 3.11.2. It is noted that calibration reference files and pipeline software are continuously updated and data fidelity is expected to improve with time, e.g., flux calibration, spike detection, treatment of 1/f-noise, etc.

Spectral extraction is done on the Level-3 data for each grating/filter combination separately using an 1.5$^{\prime\prime}$ radius sized circular aperture centered on the field-of-view, defined as an Astropy \texttt{Region}. Subsequently, the extraction aperture was converted to a pixel-weighted image mask. For each wavelength slice, missing or invalid data overlapping the aperture are linearly interpolated in the spatial domain, where the interpolation aperture is taken as a circle with a radius 3 pixels larger than that of the extraction aperture. The spectrum, in spectral units of MJy/sr, is computed as the sum of the science data cube multiplied by the image mask resized to match the cube's spectral axis and multiplied by the ratio of the area per pixel over the area of the aperture. Uncertainties are computed following the same procedure, but using the squared error and taking a final square root. In addition, the variation of the extracted signal is recorded as the standard deviation within the extraction aperture, which is a measure of the smoothness across the aperture. Lastly, Data Quality Flags are propagated by \texttt{or}-ing all flags within the extraction aperture, where interpolated pixels have bit 7 (\texttt{RESERVED}) set.

The spectrum for each grating/filter are combined and written to file in IPAC table format, tabulating wavelength, flux, flux uncertainty, flux variation, order, and flags. At the same time, a header is added that keeps track of pertinent proposal, observation, and reduction parameters using key/value pairs. Lastly, the need for stitching is assessed, which was deemed unnecessary.

Spectral extractions focusing on distinct morphological regions and features are planned for forthcoming papers.

\section{Results}
\label{sec:results}

First, the morphology of the 3.3~$\mu$m PAH emission for each target is considered. Second, the extracted spectra are explored, taking a closer look at wavelength regions of interest.

\subsection{Morphology}
\label{subsec:morphology}

\begin{figure*}
    \includegraphics[width=0.33\linewidth]{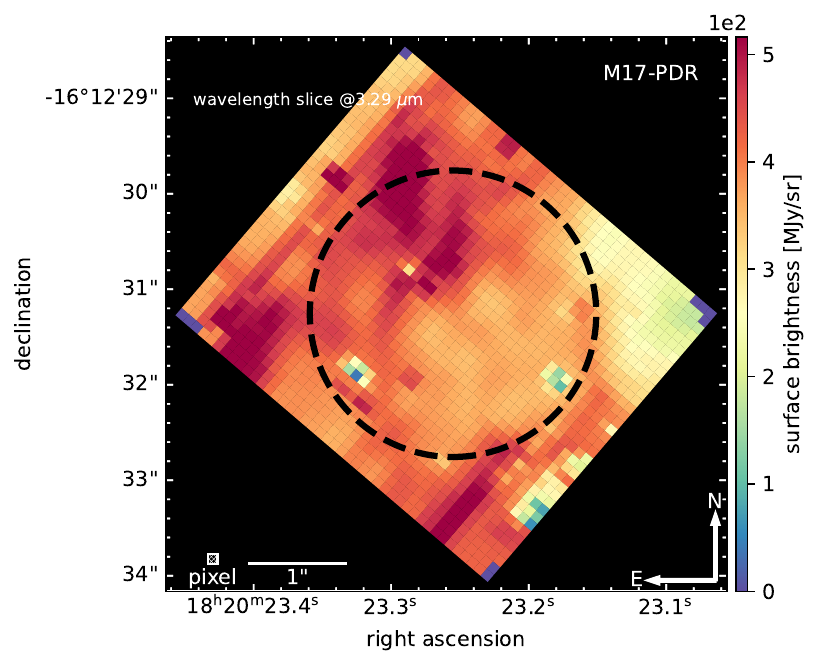}\hfill
    \includegraphics[width=0.33\linewidth]{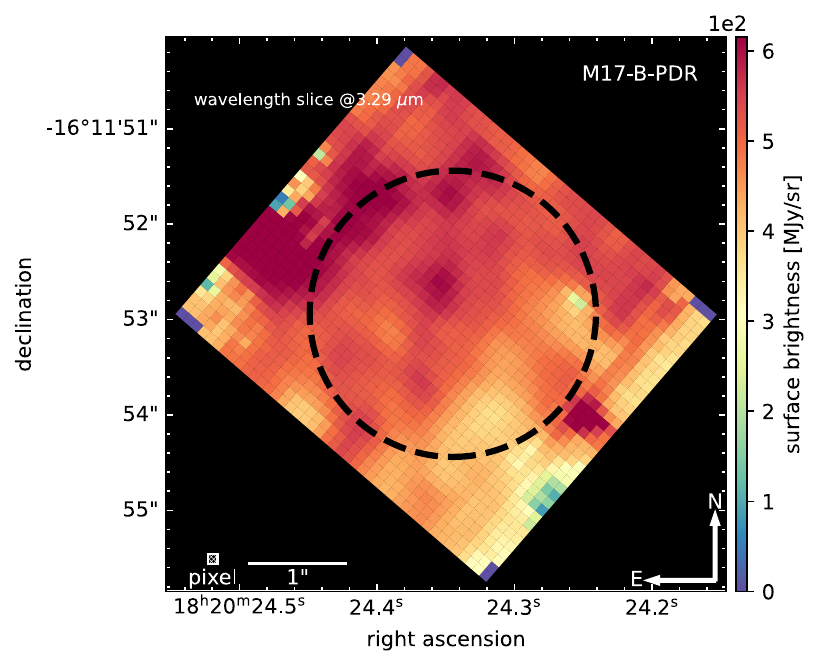}\hfill
    \includegraphics[width=0.33\linewidth]{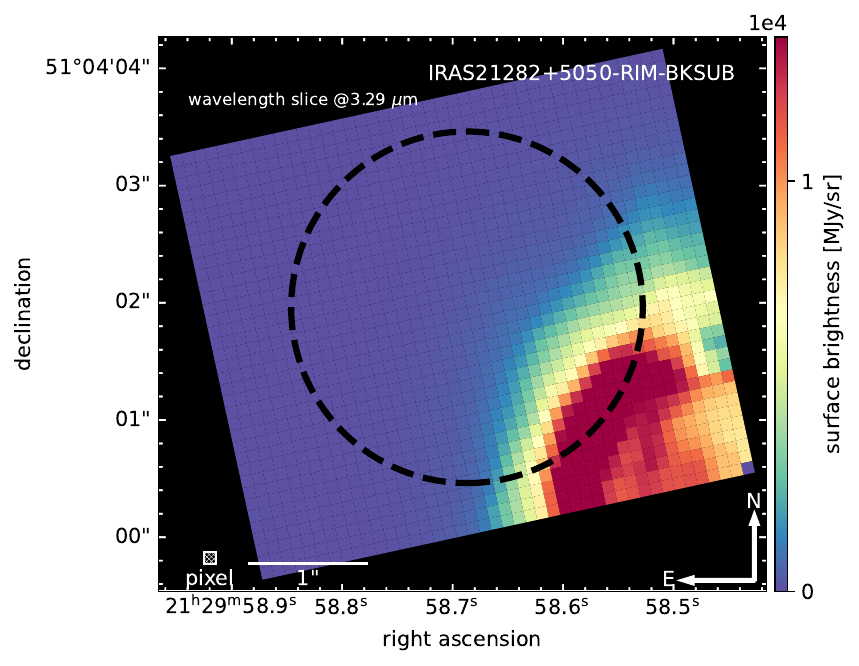}\\
    \includegraphics[width=0.33\linewidth]{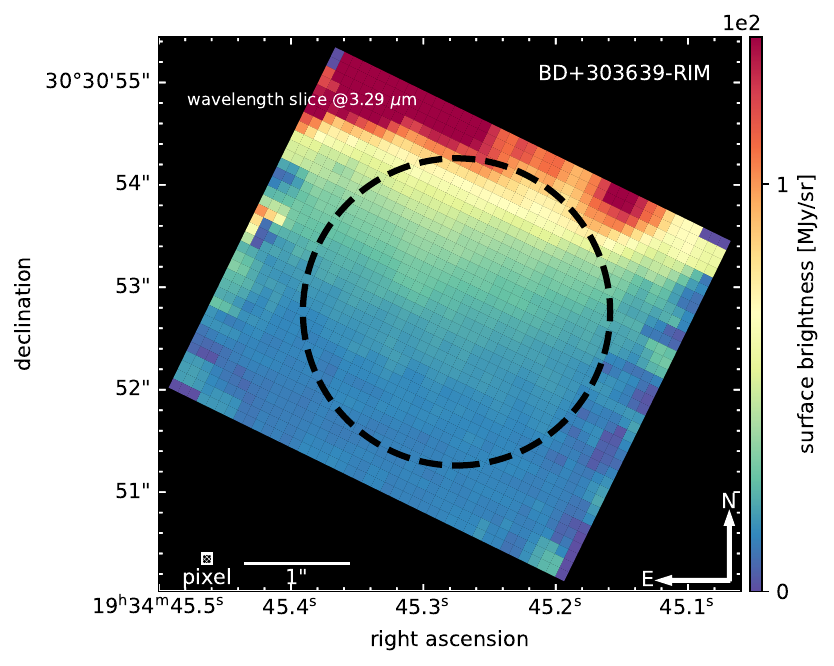}
    \includegraphics[width=0.33\linewidth]{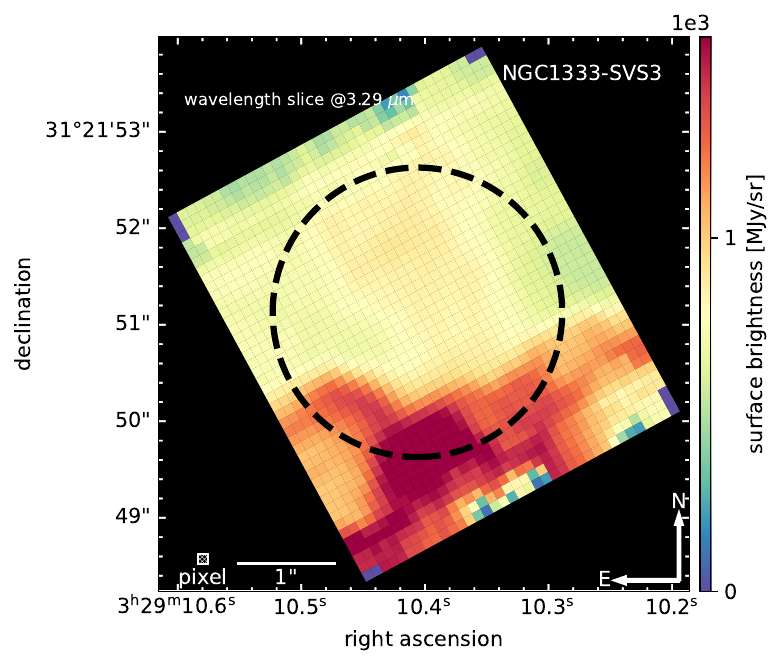}\hfill
    \includegraphics[width=0.33\linewidth]{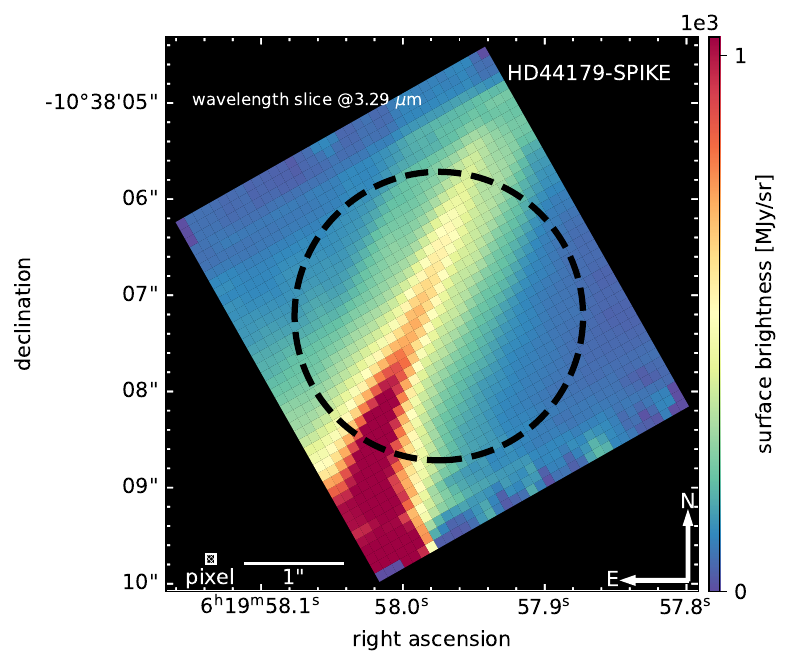}\\
    \includegraphics[width=0.33\linewidth]{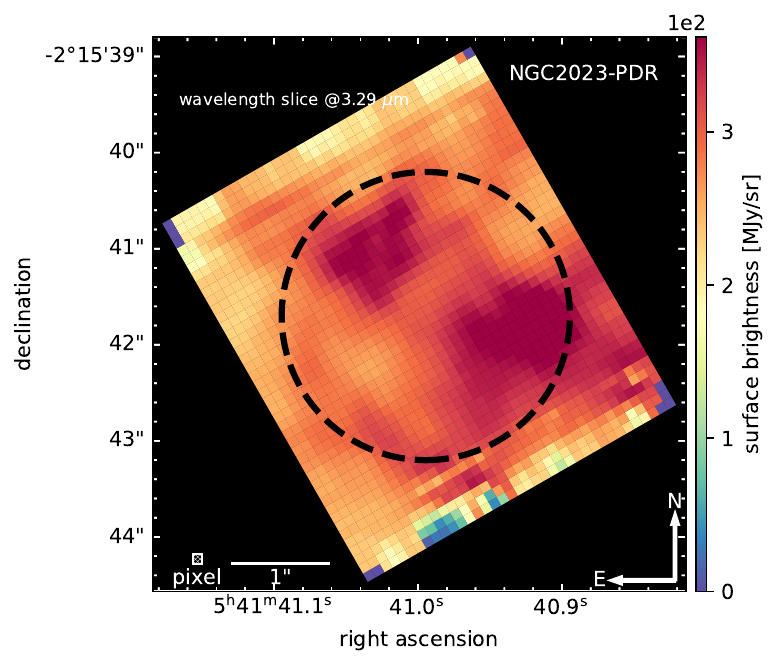}\hfill\includegraphics[width=0.33\linewidth]{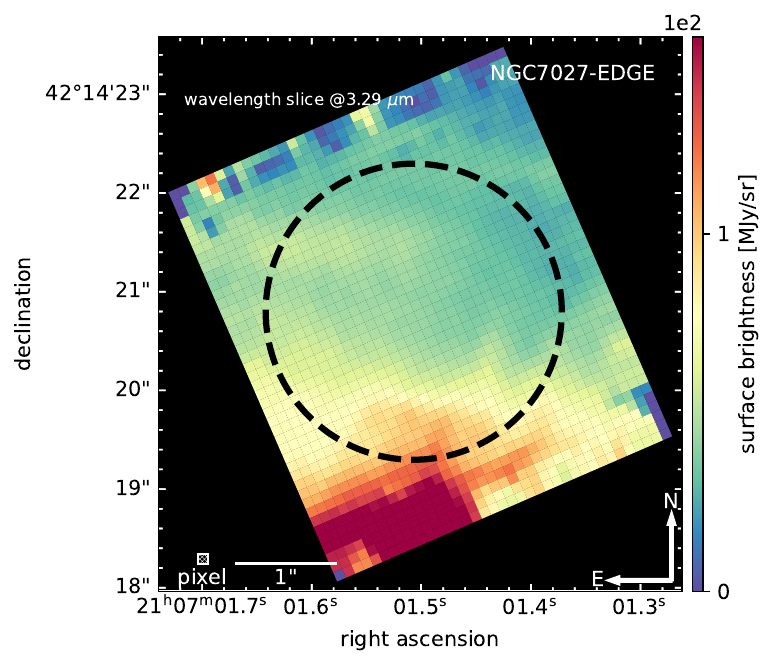}\hfill\hspace{0.33\linewidth}
    \caption{3.29~$\mu$m slices of the NIRSpec-IFU spectral-spatial data cubes. Indicated are the 1.5$^{\prime\prime}$ radius sized circular extraction apertures (black, dashed) used in this work, the $0.1^{\prime\prime}{\times}0.1^{\prime\prime}$ pixel size, a 1$^{\prime\prime}$ scale bar, and the directions of North and East.}
    \label{fig:aperture}
\end{figure*}

Figure~\ref{fig:aperture} presents the 3.29~$\mu$m slice for each target, i.e., at the peak position of the 3.3~$\mu$m PAH band. The figure shows that there is well-defined morphological substructure present in most of the images. Notably, the maps of BD+303639-RIM, IRAS21282+5050-RIM-BKSUB, HD44179-SPIKE, and NGC7027-EDGE show that their fields have been aptly labeled RIM, SPIKE and EDGE. For M17, NGC1333-SVS3, and NGC2023-PDR the substructure is less distinct and more nebulous in nature. Comparison between these maps and the imagery in Section~\ref{sec:observations} shows good agreement between the substructure observed in both. This is especially true for BD+303639-RIM, IRAS21282+5050-RIM-BKSUB, and NGC7027-EDGE, but a bit more difficult to discern for both M17 positions, NGC1333-SVS3, and NGC2023-PDR, where the field-of-view is either larger, more crowded, or both. For HD44179 there appears to be a slight offset between the 3.3~$\mu$m PAH map and the image that is larger than the 0.06$^{\prime\prime}$ reported rms in ra and dec for the Hubble data. It is noted that the Hubble Legacy Archive offers a number of observations in the same filters having differing coordinate solutions; those used here match the NIRSpec observations best.

\subsection{The Spectra}
\label{subsec:spectra}

\begin{figure*}
    \includegraphics[width=0.33\linewidth]{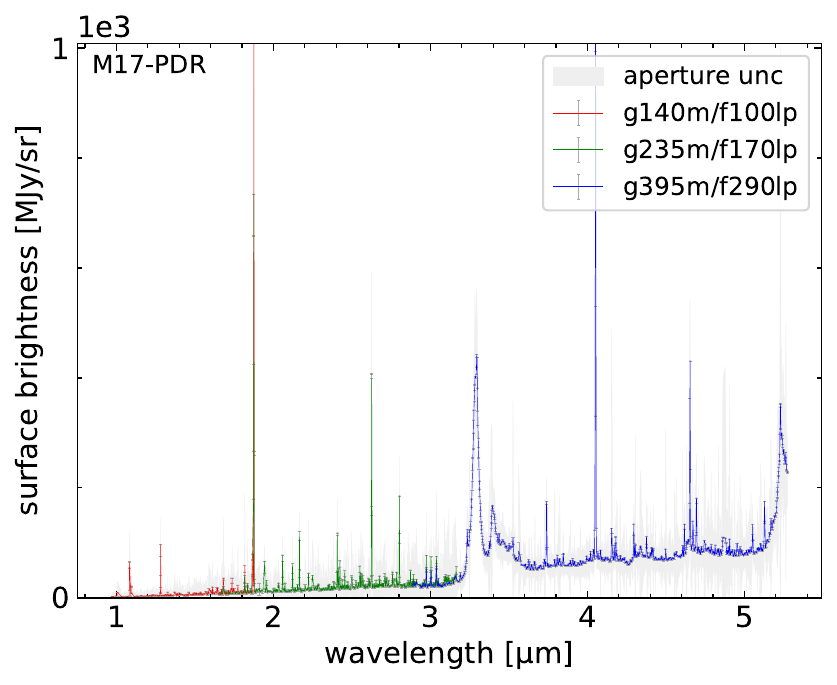}\hfill
    \includegraphics[width=0.33\linewidth]{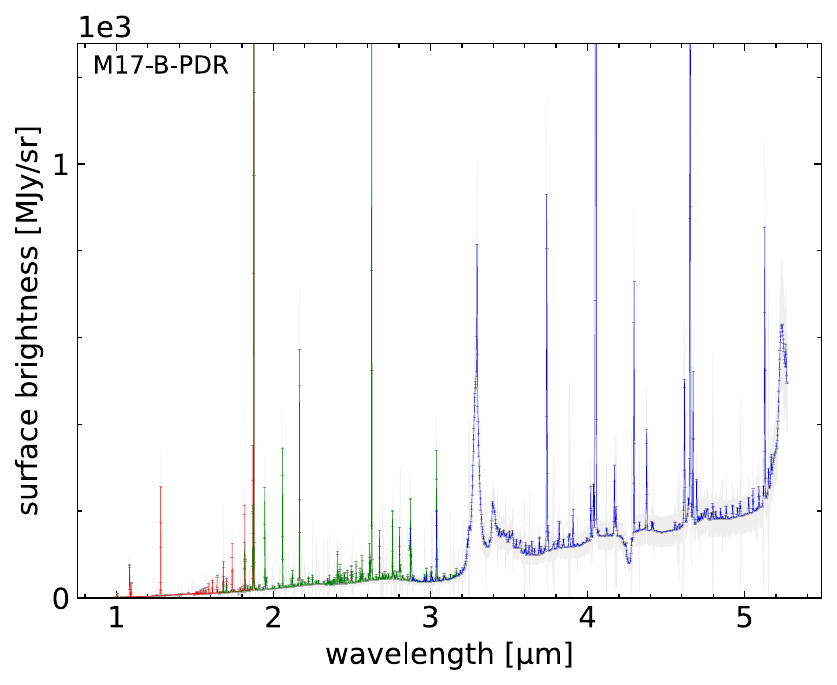}\hfill
    \includegraphics[width=0.33\linewidth]{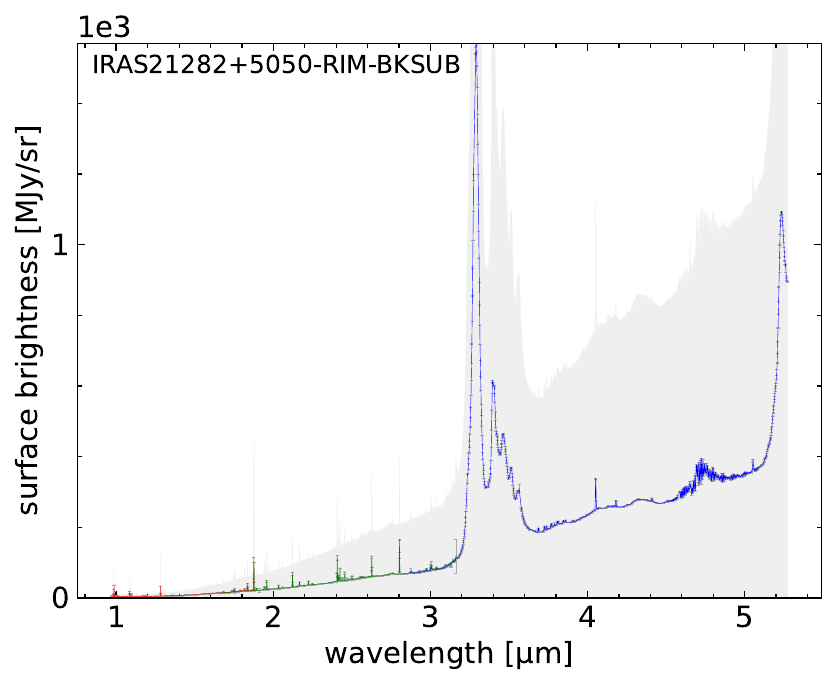}\\
    \includegraphics[width=0.33\linewidth]{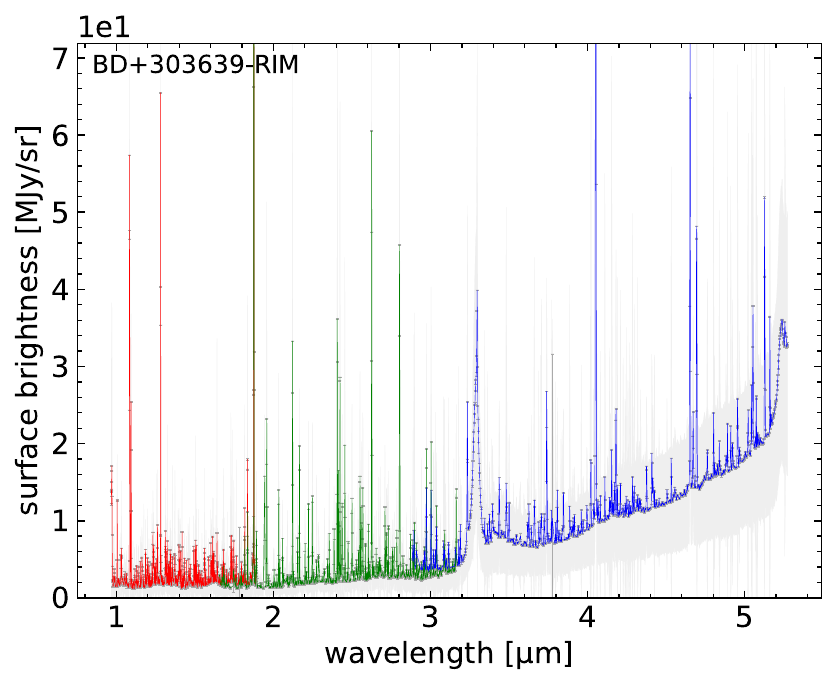}\hfill
    \includegraphics[width=0.33\linewidth]{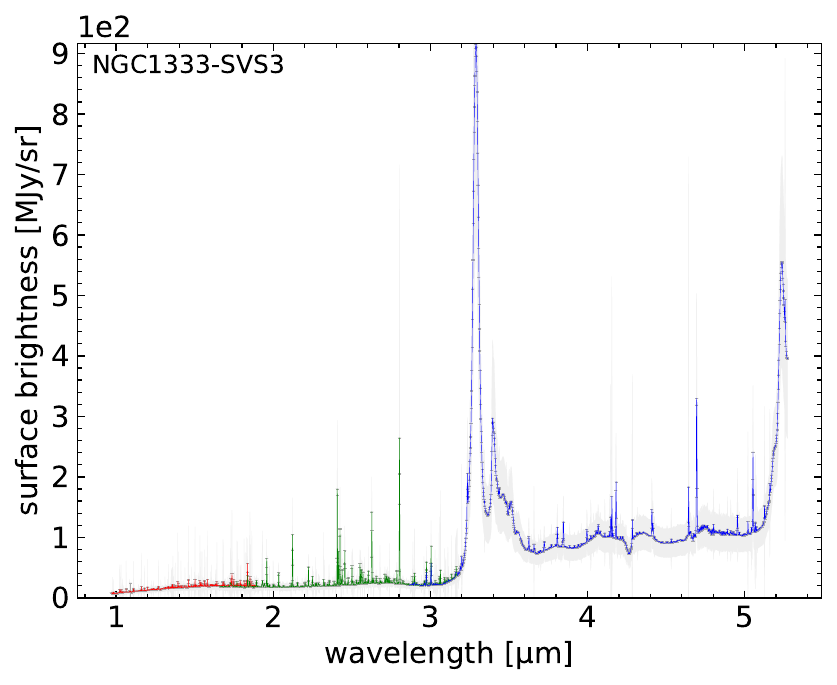}\hfill
    \includegraphics[width=0.33\linewidth]{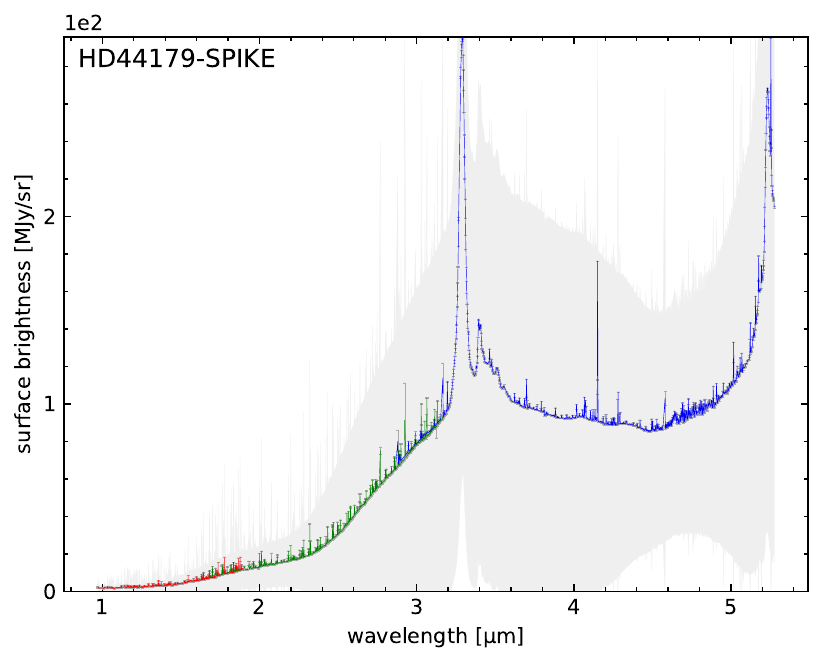}\\
    \includegraphics[width=0.33\linewidth]{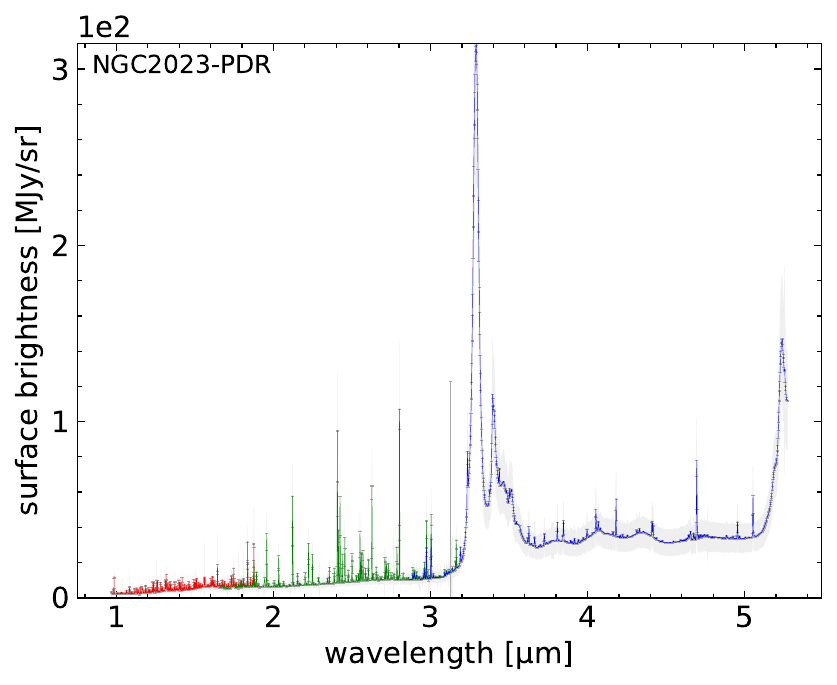}\hfill\includegraphics[width=0.33\linewidth]{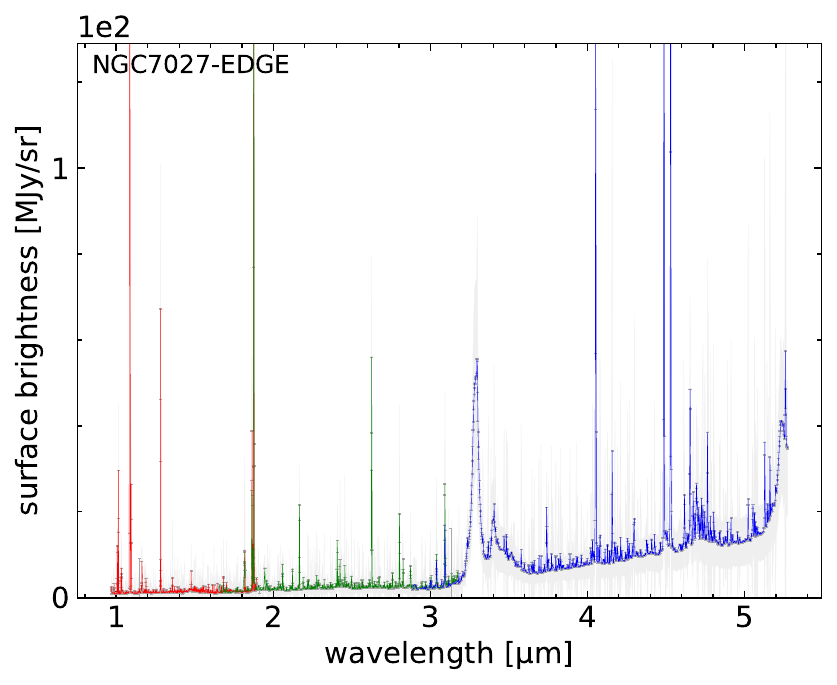}\hfill\hspace{0.33\linewidth}
    \caption{NIRSpec-IFU 1-5~$\mu$m spectra for each target extracted from a 1.5$^{\prime\prime}$ radius sized circular aperture. Each grating/filter combination has been color-coded and 1-$\sigma$ error bars are indicated (darker grey), which are typically smaller than the line width. The envelopes (grey) show the 1-$\sigma$ variation of the spectral signal within the extraction aperture.}
    \label{fig:spectra}    
\end{figure*}

Figure~\ref{fig:spectra} presents the spectra extracted for each target from their associated 1.5$^{\prime\prime}$ radius sized circular aperture depicted in Figure~\ref{fig:aperture}. The spectra have been presented such as to capture most of their salient features. With the exception of HD44179-SPIKE, the spectra show a rising continuum with wavelength that steps up around 3.2~$\mu$m. The spectrum of HD44179-SPIKE is unusual in that the continuum rises up to the 3.3~$\mu$m PAH band, after which it drops again. Investigating this further revealed that the diffraction spike from a secondary mirror strut is aligned with the leg of the `x' in the NIRSpec field-of-view, contaminating the extracted spectrum (priv. comm. \textit{JWST} Help Desk).

All spectra have an easily distinguishable, strong 3.3~$\mu$m PAH band, a discernible 3.4~$\mu$m feature, a sharp rise at the onset of the 5.25~$\mu$m PAH band, and numerous emission lines. Notably, BD+303639-RIM shows a relatively strong forest of emission lines, whereas both M17 positions, NGC1333-SVS3, NGC2023-PDR, and NGC7027-EDGE have a plethora of relatively weaker lines. For IRAS21282+5050-RIM-BKSUB and HD44179-SPIKE the emission lines are more subdued. The spectra of the two M17 positions and NGC1333-SVS3 show CO$_{\rm 2}$-ice absorption around 4.27~$\mu$m and signs of a very broad H$_{\rm 2}$O-ice absorption band, most apparent between 2.8-3.2~$\mu$m. The spectra of IRAS21282+5050-RIM, HD44179-SPIKE, NGC1333-SVS3, and NGC7027-EDGE show signs of CO gas emission around 4.66~$\mu$m.

The variation of the spectral signal within the extraction aperture (grey envelopes) differs considerably between the targets. It is most notable for IRAS21282-RIM and HD44179-SPIKE and is somewhat present for BD+303639-RIM and NGC7027-EDGE. This is consistent with the morphology of these sources depicted in Figure~\ref{fig:aperture}. That is, they either have a strong gradient in surface brightness or a distinct morphological feature within the extraction aperture.

The PAH-related features in the spectra are explored in the next section, followed by the CO$_{\rm 2}$-ice absorption band, CO gas, atomic, and molecular emission lines. Besides a direct spectral comparison between targets, some astrophysical analysis is performed to provide a flavor of the kind of information that is contained in the data. Since the spectra have been extracted from a single, large aperture that for most targets includes edges, rims, ridges, etc., the analysis is predominantly limited to M17-B-PDR. In-depth spectral-spatial analysis of \emph{all} features and targets are planned for a series of forthcoming papers, as it is beyond the scope of this first, exploratory, look.

\subsubsection{PAH Spectral Features}
\label{subsubsec:PAHFeatures}

PAH-related spectral regions of interest are considered next.

\begin{figure}
    \includegraphics[width=\linewidth]{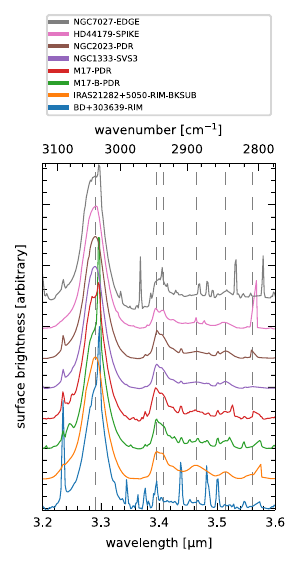}
    \caption{3.2-3.6~$\mu$m region for each target, continuum subtracted, normalized to the peak strength of the 3.3~$\mu$m PAH feature, and offset. The grey dashed lines indicate positions of the broad PAH-related features. See Section~\ref{subsubsec:pah33} for a discussion.}
    \label{fig:pah33}
\end{figure}

\begin{figure}
    \centering
    \includegraphics[width=\linewidth]{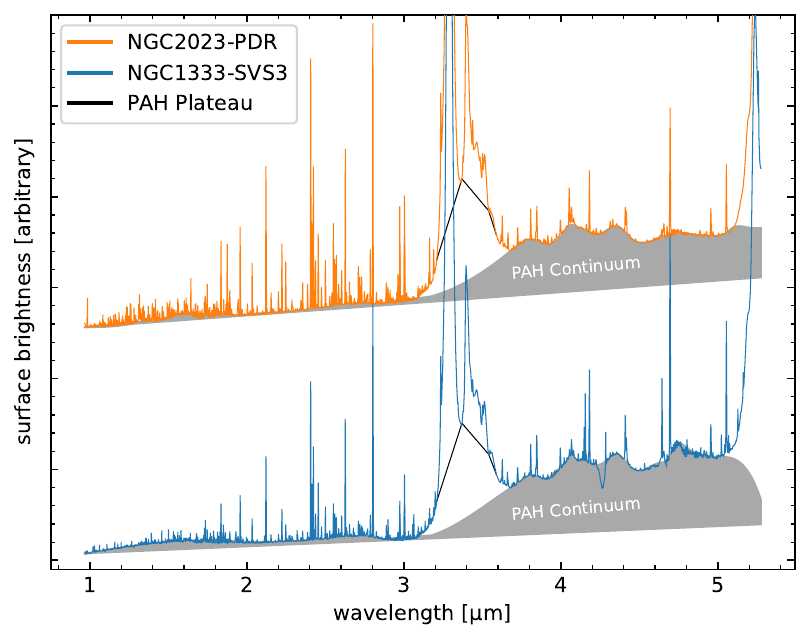}
    \caption{1-5~$\mu$m spectra of NGC2023-PDR and NGC1333-SVS3 indicating the PAH-continuum (dark grey). The triangular shaped region below the 3.3/3.4~$\mu$m band complex (black line segments) designates emission associated with the 3~$\mu$m band complex. See Section~\ref{subsubsec:quasi} for a discussion.}
    \label{fig:quasi-continuum}
\end{figure}

\paragraph{The 3.2-3.6~$\mu$m, CH stretch region}Figure~\ref{fig:pah33} compares the 3.2-3.6~$\mu$m region for each target. To separate the broad PAH-related features from the underlying plateau, a local spline continuum has been subtracted with knots at the fixed wavelength positions of 3.21, 3.37, 3.54, and 3.59~$\mu$m (see also Figure~\ref{fig:quasi-continuum}). Besides the dominant 3.3~$\mu$m band with its blue shoulder, distinct, broad features peaking roughly at 3.395, 3.408, 3.465, 3.515, and 3.560 ~$\mu$m are common to all spectra. The 3.3~$\mu$m PAH band peaks at 3.29~$\mu$m for most targets, with a few peaking at slightly longer wavelengths due to the H\textsc{i} Pf:9-5 recombination line being superimposed. The relative intensities of the PAH-related features show some variation between objects. The strongest of these, centered at 3.4~$\mu$m, is a doublet with separate components at 3.395 and 3.408~$\mu$m. The 3.408 and 3.465~$\mu$m bands are most prominent in IRAS21282+5050-RIM-BKSUB.

\begin{figure}
    \includegraphics[width=\linewidth]{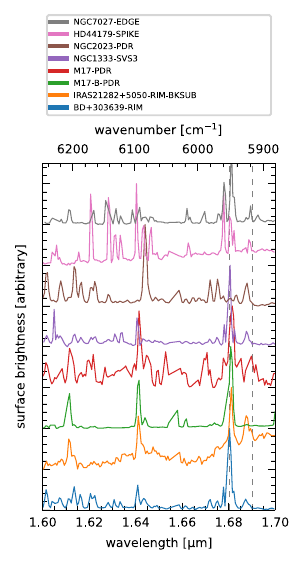}
    \caption{1.6-1.7$~\mu$m region for each target, scaled and offset. The grey dashed lines bracket the nominal position of the $v$=2-0 overtone of the 3.3~$\mu$m PAH band. It is noted that the triangular-like structure seen between $\sim$1.655-1.659~$\mu$m is the result from an incomplete removal of a data spike. See Section~\ref{subsubsec:overtone} for a discussion.}
    \label{fig:overtone}
\end{figure}

\paragraph{The 1.6-1.7~$\mu$m, CH stretch overtone region}The first overtone of the 3.3~$\mu$m PAH band is predicted to fall in this region \citep[]{1994ApJ...429L..91D, 2019A&A...632A..71C}, with a tentative detection reported by \citet{1994ApJ...434L..15G} in IRAS21282+5050. Figure~\ref{fig:overtone} zooms in on the 1.6-1.7~$\mu$m region region for each target. The spectra have been scaled by setting the minimum in the region to zero and the height to half that of the strongest line. Except perhaps for IRAS21282+5050-RIM-BKSUB, the figure shows that there is little evidence for distinct emission near 1.68~$\mu$m that can directly be associated with an overtone of the 3.3~$\mu$m PAH band. For IRAS21282+5050-RIM-BKSUB, a broad undulating ``continuum'' can be seen that rises and falls from 1.63 to 1.67~$\mu$m and rises again from 1.67 to longwards of 1.7~$\mu$m.

\begin{figure}
    \includegraphics[width=\linewidth]{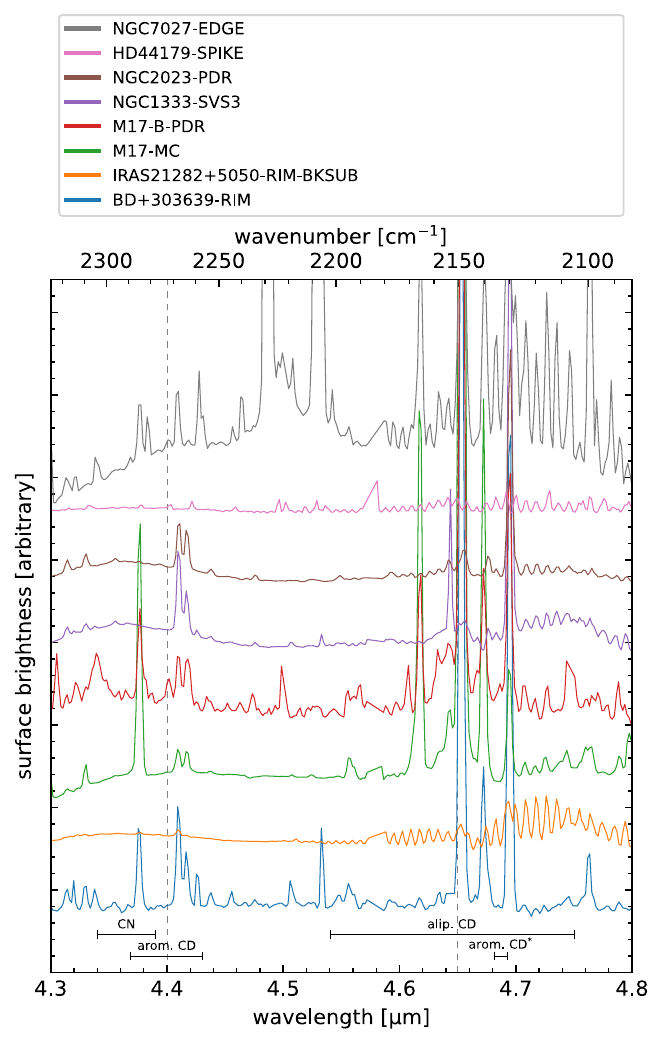}
    \caption{4.3-4.8~$\mu$m region for each target, continuum subtracted and offset. Indicated are predicted regions where the aromatic and aliphatic CD stretches and the CN stretch in PAH-nitriles fall (\citealt{1997JPCA..101.2414B}; \citealt{2004ApJ...614..770H}; \citealt{2021ApJ...917L..35A}; Esposito~et~al. in prep.). The grey dashed lines indicate their typical astronomical positions. $^{\rm *}$Range found for the aromatic CD stretch in pentacene. See Section~\ref{subsubsec:d+cn} for a discussion.}
    \label{fig:d+cn}
\end{figure}

\paragraph{The 4.3-4.8~$\mu$m Deuterated and PAH-nitriles Region}Figure~\ref{fig:d+cn} compares the 4.3-4.8~$\mu$m region for each target, indicating where emission features from deuterated- and PAH-nitriles have been predicted to fall \citep[e.g.,][and references therein]{2020ApJ...892...11B, 2021ApJ...917L..35A}. For the figure, each spectrum has been continuum subtracted using a spline with anchor points at 4.30, 4.45, 4.60, and 4.78~$\mu$m, and is offset. Except for NGC7027-EDGE, this removes the continuum adequately, leaving only a small amount of residual broad band structure in a few spectra.

Any clear detection of the \textit{aliphatic} CD stretch at 4.65~$\mu$m is severely hampered by CO emission in the spectra of HD44179-SPIKE, IRAS21282+5050-RIM-BKSUB, NGC1333-SVS3, and NGC7027-EDGE. For M17-PDR, M17-B-PDR, and NGC1333-SVS3 such a feature could be hidden in the broad blended emission complex apparent in their spectra. The remaining spectra show no clear evidence for this feature. Similarly, the \textit{aromatic} CD band centered around 4.4~$\mu$m is also confused by line emission.

\begin{figure*}
    \includegraphics[width=0.5\linewidth]{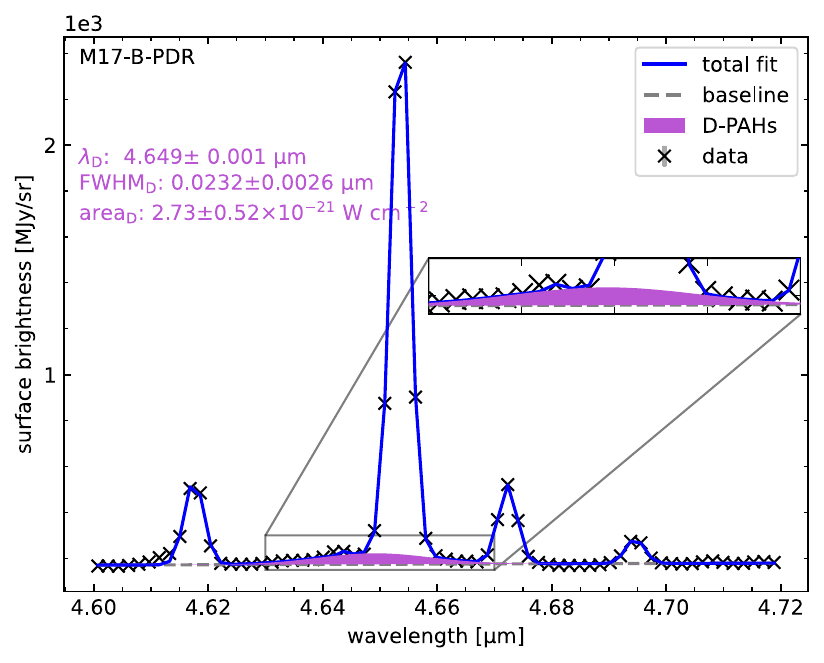}\hfill
    \includegraphics[width=0.5\linewidth]{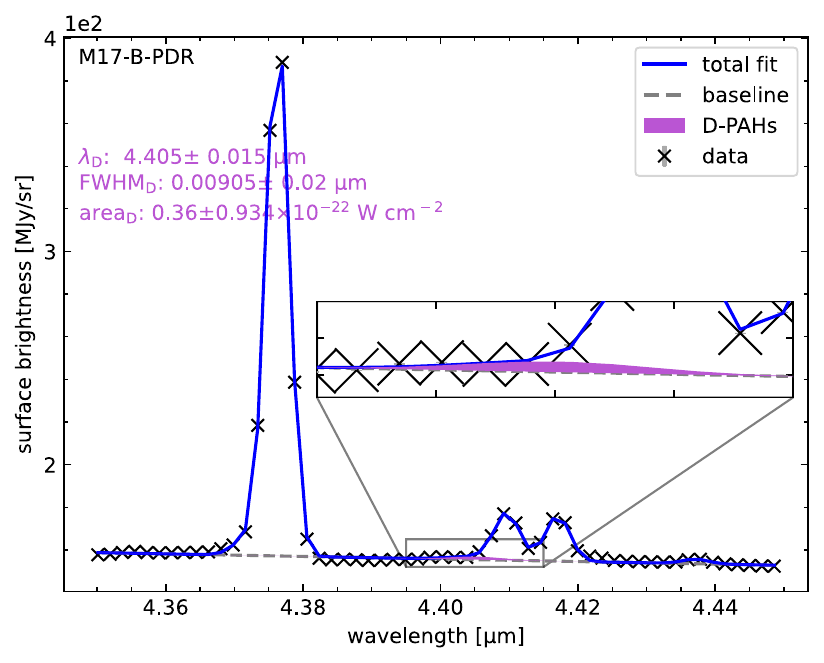}
    \caption{Zoom-in and multi-component decomposition of the 4.60-4.72~$\mu$m (aliphatic region; left) and 4.33-4.47~$\mu$m (aromatic region; right) spectrum of M17-B-PDR. The properties of the component possibly attributable to deuterated PAHs (purple) are given. See Section~\ref{subsubsec:d+cn} for a discussion.}
    \label{fig:ali+aro}
\end{figure*}

Following the approach by \citet{2016A&A...586A..65D}, the left panel of Figure~\ref{fig:ali+aro} decomposes the 4.60-4.72~$\mu$m spectrum of M17-B-PDR into multiple components by simultaneously fitting a 1$^{\rm st}$ order polynomial baseline, 6 narrow Gaussian profiles to match the emission lines, and 1 broad Gaussian representing emission that may be attributable to the \textit{aliphatic} CD stretch. A feature is recovered at 4.649$\pm$0.001~$\mu$m, consistent with the astronomical position reported by \citet{2016A&A...586A..65D}. The feature has an integrated intensity of 2.73$\pm$0.52$\times$10$^{\rm -21}$~W~cm$^{\rm -2}$ and a signal-to-noise-ratio of 5.25.

Likewise, the right panel of Figure~\ref{fig:ali+aro} decomposes the 4.35-4.45~$\mu$m spectrum of M17-B-PDR using a similar approach to search for emission attributable to the \textit{aromatic} CD stretch. However, there is no compelling evidence for such a feature (signal-to-noise-ratio of 0.39).

Combining the deuterated PAH intensities with values determined for the 3.3~$\mu$m aromatic CH stretch and 3.4~$\mu$m aliphatic CH stretch bands in M17-B-PDR of 7.28$\pm$0.493$\times$10$^{\rm -20}$ and 7.53$\pm$1.48$\times$10$^{\rm -21}$~W~cm$^{\rm -2}$, respectively, yields I$_{\rm 4.4}$/I$_{\rm 3.3}$=4.94$\pm$12.83$\times$10$^{\rm -4}$ and I$_{\rm 4.65}$/I$_{\rm 3.4}$=0.36$\pm$0.100.

Then, for simplicity, assuming that the integrated cross sections for corresponding aromatic and aliphatic C-H/C-D modes are comparable \citep[][]{2004ApJ...604..252P} and following \citet{2020ApJS..251...12Y}, the degree of deuteration can be estimated. Here, the ratio of integrated cross sections A$_{\rm 4.40}$/A$_{\rm 3.3}$=0.56$\pm$0.19 and the blackbody intensity ratio B(3.3)/B(4.4)=0.70$\pm$0.28 (400${\lesssim}T_{\rm PAH}{\lesssim}$900~K). With the observed ratio I$_{\rm 4.4}$/I$_{\rm 3.3}$, the degree of \emph{aromatic} deuteration [D/H]$_{\rm arom.}$=0.06$\pm$0.16\%. Similarly, with the observed \emph{aliphatic} ratio I$_{\rm 4.65}$/I$_{\rm 3.4}$, [D/H]$_{\rm alip.}$=31$\pm$12.7\% for M17-B-PDR.

\paragraph{The 1-5~$\mu$m PAH Continuum}Figure~\ref{fig:quasi-continuum} indicates what is dubbed here the PAH-continuum \citep[initially called the ``the vibrational quasi-continuum'';][]{1989ApJS...71..733A}. The 1-5~$\mu$m portion starts near 1.2~$\mu$m, rises slowly to $\sim$3.2~$\mu$m, where it steps up by a factor of $\sim$2.5, after which it stays relatively flat out to 5~$\mu$m. This step-up is distinct from the emission underlying the 3.3/3.4~$\mu$m PAH band complex indicated in the figure. Broad bumps that are part of this continuum are centered near 3.8, 4.04, and 4.34~$\mu$m can be seen in all targets but BD+303639+RIM and NGC7027-EDGE, where they are possibly hidden by the forest of intense emission lines.

\subsubsection{CO\texorpdfstring{\textsubscript{2}}{2}/H\texorpdfstring{\textsubscript{2}}{2}O-ice Absorption, CO Gas, and Line Emission}
\label{subsubsec:other}

The spectra in Figure~\ref{fig:spectra} show signs of CO$_{\rm 2}$- and H$_{\rm 2}$O-ice absorption bands, gas-phase CO ro-vibrational lines, and atomic and molecular line emission, which are considered next.

\begin{figure*}
    \includegraphics[width=0.5\linewidth]{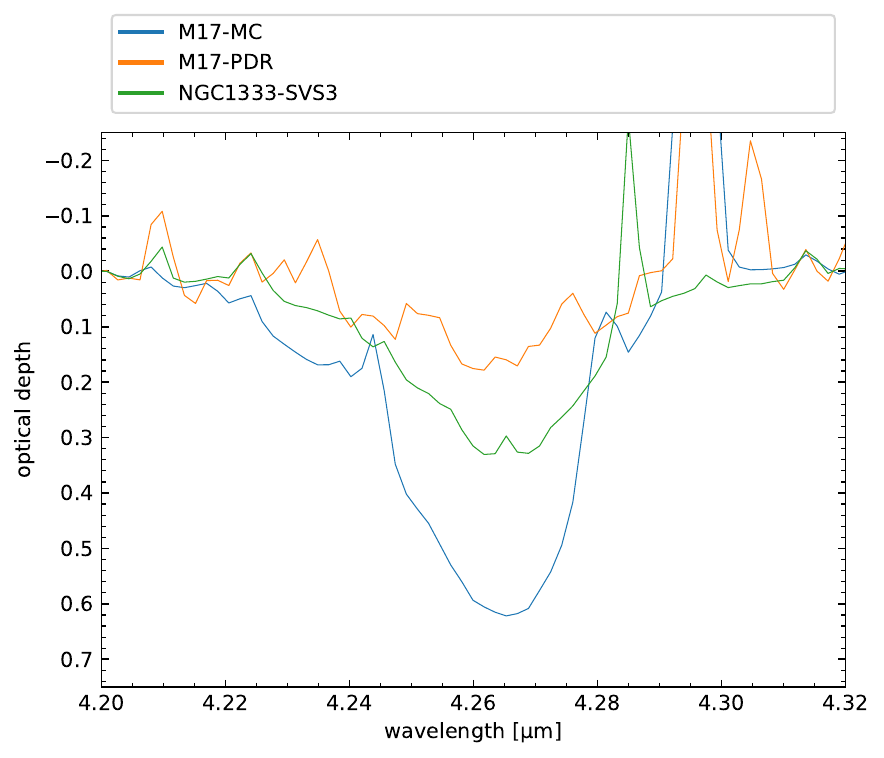}\hfill\includegraphics[width=0.5\linewidth]{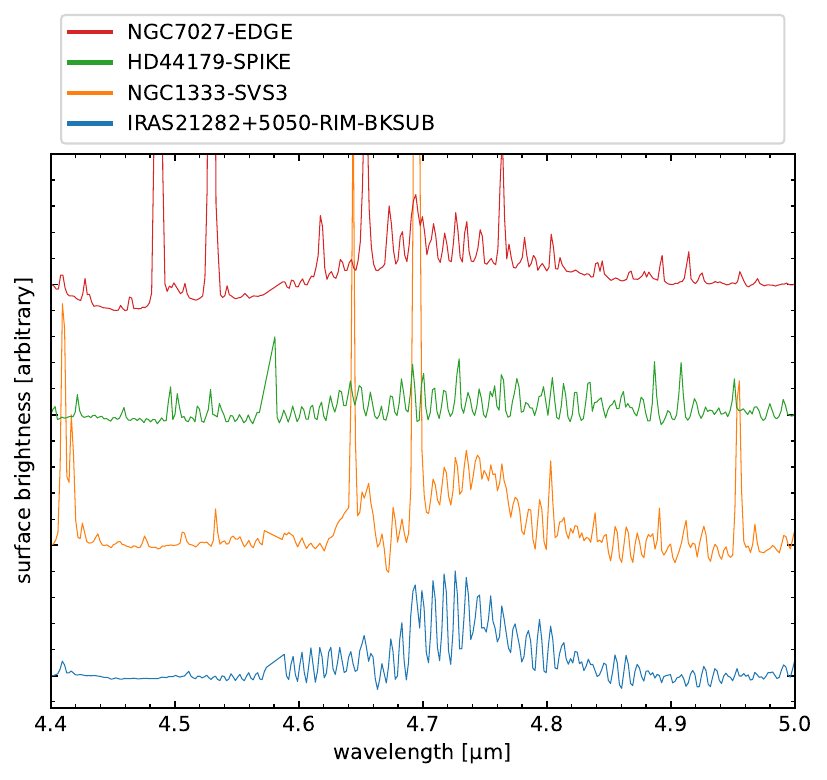}
    \caption{Comparison of the observed 4.2~$\mu$m CO$_{\rm 2}$ ice absorption (left) and CO ro-vibrational emission (right). See Sections~\ref{subsec:ices} and \ref{subsec:lines} for discussions.}
    \label{fig:sections}
\end{figure*}

\paragraph{The 4.2-4.3~$\mu$m CO\textsubscript{2}- and 2.8-3.2~$\mu$m H$_{\rm 2}$O-Ice Absorption Bands}The left panel of Figure~\ref{fig:sections} compares the 4.2~$\mu$m CO$_{\rm 2}$ ice absorption signature detected at both M17 positions and NGC1333-SVS3. A spline with anchor points near 4.200, 4.318, and 4.350~$\mu$m has been used to determine the optical depth of each of the CO$_{\rm 2}$-ice features. The figure shows quite some variation, with maximum optical depths of $\sim$0.6, 0.3, 0.1 for M17-B-PDR, M17-PDR, and NGC1333-SVS3, respectively.

\begin{figure}
    \includegraphics[width=\linewidth]{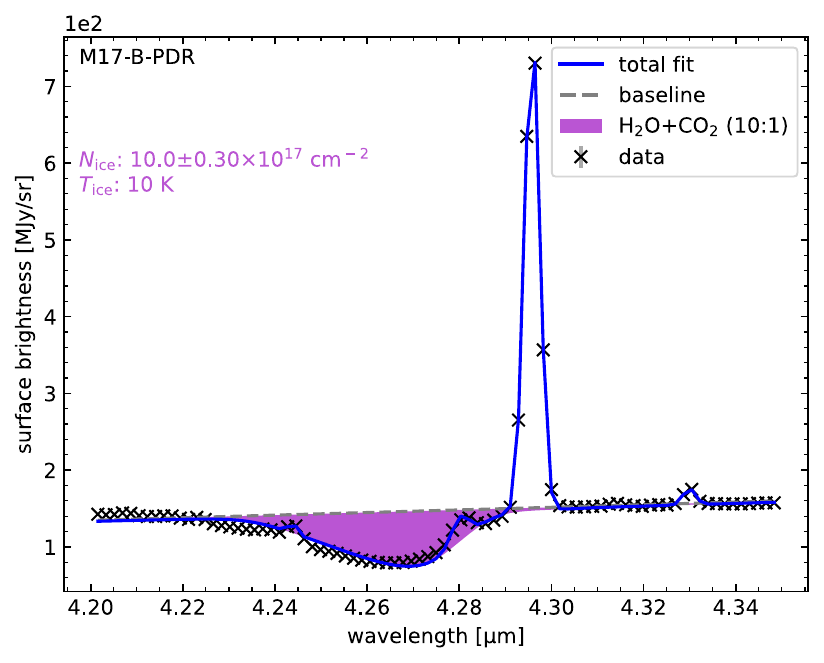}
    \caption{Multi-component fit of the CO\textsubscript{2} ice absorption band in the spectrum of M17-B-PDR. The shaded area (purple) indicates the H$_{\rm 2}$:CO$_{\rm 2}$ (10:1), 10~K ice component. See Section~\ref{subsec:ices} for a discussion.}
    \label{fig:co2}
\end{figure}

Utilizing the absorption data for an H$_{\rm 2}$O+CO$_{\rm 2}$ (10:1) ice at 10~K from the Leiden Ice Database \citep[][]{1999A&A...350..240E, 2022A&A...668A..63R}, Figure~\ref{fig:co2} shows the result of a simultaneous fit to the 4.20-4.35~$\mu$m spectrum of M17-B-PDR with a 1$^{\rm st}$ order polynomial baseline, 4 Gaussian profiles for the emission lines and attenuation by absorption of the CO$_{\rm 2}$ water-ice. This produces a good match, yielding a column density of 10.0$\pm$0.30$\times10^{\rm 17}$~cm$^{\rm -2}$ for the ice.

Lastly, the very broad and strong H$_{\rm 2}$O-ice absorption band falling between 2.8-3.2~$\mu$m is observed in the spectra of M17-PDR, M17-B-PDR, and NGC1333-SVS3 (see Figure~\ref{fig:spectra}). Though, its analysis is more challenging due to the overlap with the 3.3~$\mu$m PAH complex.

\paragraph{The 4.6-4.9~$\mu$m CO Gas Emission Lines}The right panel of Figure~\ref{fig:sections} compares the ro-vibrational structure of CO gas emission that is apparent in the spectra of IRAS21282+5050-RIM-BKSUB, HD44179-SPIKE, NGC1333-SVS3, and NGC7027-EDGE. Here, a local spline continuum with anchor points at 4.400, 4.500, 4.605, 4.900, and 4.995~$\mu$m is first subtracted from each spectrum, after which it is scaled to emphasize the ro-vibrational structure.

\begin{figure}
    \includegraphics[width=\linewidth]{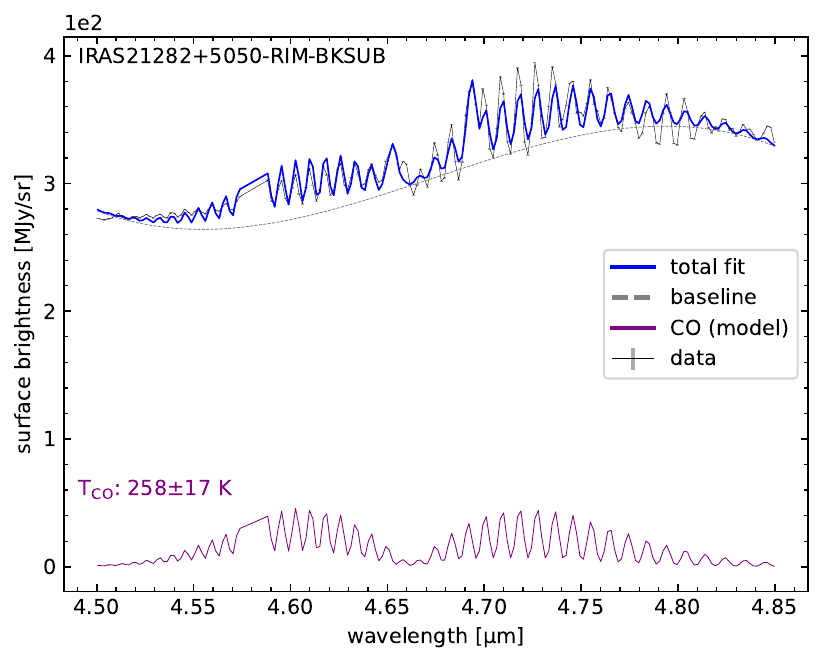}
    \caption{Warm, 258~K, CO gas detected in emission in the spectrum of IRAS21282+5050-RIM-BKSUB. Note that the flat appearing structure around 4.58~$\mu$m is due to a spike in the data being masked. See Section~\ref{subsec:lines} for a discussion.}
    \label{fig:co}
\end{figure}

Figure~\ref{fig:co} simultaneously fits the 4.50-4.85~$\mu$m spectrum of IRAS21282+5050-RIM-BKSUB with a third order polynomial, Gaussian emission lines at 4.65 and 4.70~$\mu$m, and a simple rigid-rotor CO model with, as free parameters, the power and temperature. The fit is very reasonable, especially given the simplicity of the employed CO model, and indicates warm CO gas at a temperature of $T_{\rm CO}$=258$\pm$17~K.\\

\paragraph{Emission Lines}The spectra in Figure~\ref{fig:spectra} show a plethora of emission lines, spread across the entire 1-5~$\mu$m region. To generate an inventory, an uncurated list of some 17,000 lines between 0.6-5.3~$\mu$m is considered. This list contains transition data for atomic Al, Ar, C, Ca, Co, Fe, He, H I, K, Mg, N, Ni, O, P, and molecular H$_{\rm 2}$ and HD. Here, the line list for H$_{\rm 2}$ from \citet{2019A&A...630A..58R} and the line list for HD from HITRAN \citep[][]{2021NatRP...3..302R} are combined with atomic data from the Atomic Line List\footnote{\url{www.pa.uky.edu/~peter/newpage/index.html}} \citep[v3.00b4;][]{2018Galax...6...63V}. Each line is blindly fitted to the spectrum, per grating/filter combination, in $\lambda$-$F_{\lambda}$-space, using a Gaussian line profile together with a 1$^{\rm st}$-order polynomial baseline. The fit is performed across the wavelength range made up by 14 adjacent resolution elements around the line center, allows for up to $\pm10^{\rm -4}$~$\mu$m variation in the line position and a minimum and maximum FWHM of 0.001 and 0.02~$\mu$m, respectively. Power and line width are forced strictly positive. Line strength, propagated uncertainty and contrast are determined from each fit, where lines with a signal-to-noise ratio greater than 2, a contrast above 5\%, and an integrated intensity of at least 10$^{\rm -23}$~W~cm$^{\rm -2}$ are recorded.

\begin{figure*}
    \includegraphics[width=\linewidth]{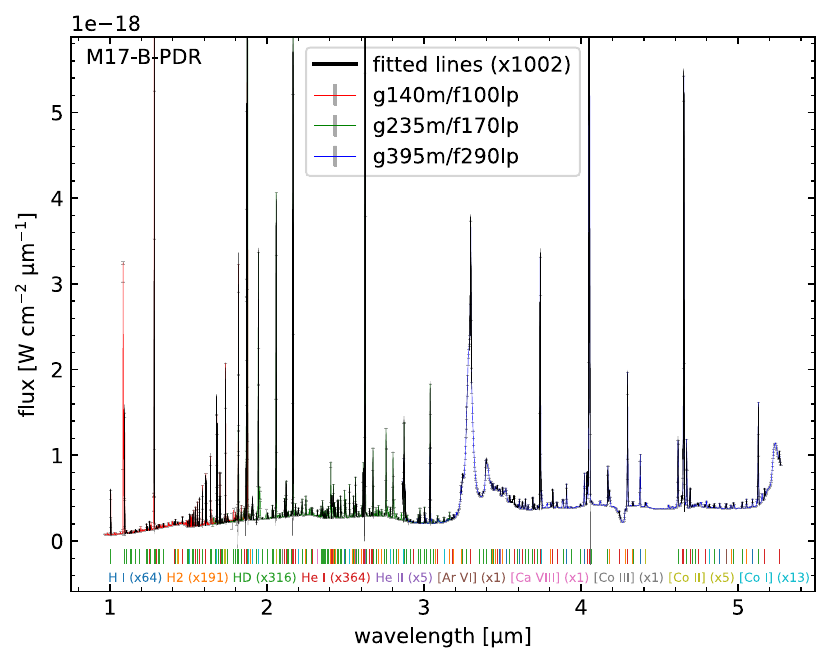}
    \caption{Line analysis of M17-MC, given a line list of some 17,000 lines results in 1002 matches. Only lines with a signal-to-noise ratio of at least 2, a contrast of 5\%, and an integrated intensity of at least 10$^{\rm -23}$~W~cm$^{\rm -2}$ are shown. Each species/ionization state has been color-coded separately. The number of lines matched per species is provided in parenthesis. Note that not all lines are matched and that there are those that are simultaneously matched by several species. Only 198 lines are spaced apart more than 5$\times10^{\rm -3}\ \mu$m.}
    \label{fig:lines}
\end{figure*}

Figure~\ref{fig:lines} presents the results for M17-B-PDR, where 1002 lines from a total of 7 species have been matched. Still, a number of unmatched lines remain. There is clearly a considerable amount of line confusion. That is, there are only 198 lines that are spaced more than 5$\times10^{\rm -3}\ \mu$m apart. This confusion is driven by an enormous amount of closely spaced He and HD lines, and not applying any priors based on the astrophysical environment to down-select the line list. For example, the
match of high-energy lines such as [Ar VI] and [Ca VIII] are hard to reconcile with the astrophysical environment of M17-B-PDR. Similarly, the presence of cobalt is highly unlikely given its low Cosmic abundance and that no matching lines were found for other metals, e.g., iron.\\

\begin{figure*}
    \includegraphics[width=\linewidth]{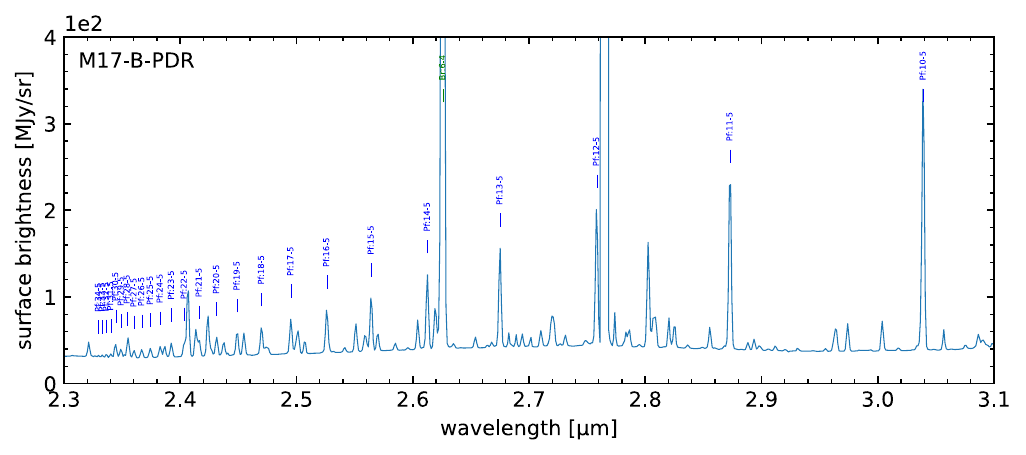}
    \caption{The 2.3-3.1~$\mu$m spectrum of M17-B-PDR depicting a rich inventory of HI emission lines. Notably, the Pfund series, where each transition has been labeled separately.}
    \label{fig:hi}
\end{figure*}

\noindent\underline{HI-lines}: Further demonstrating the richness and fidelity of the line data, Figure~\ref{fig:hi} zooms in on the atomic hydrogen Pfund series, starting from an upper level of 10 to well into the 30-ies. These lines can be, e.g., a powerful diagnostic for the geometry of the source \citep[][]{2002A&A...386L...5L}.

HI recombination lines can be used to determine the extinction along the line of sight \citep[][]{Gurzadyan1997}. For example, considering the Br$\beta$ and Br$\alpha$ lines at 2.63 and 4.05~$\mu$m, respectively, and case B recombination \citep[][]{1938ApJ....88...52B}, the total visual extinction $A_{\rm V}$ can be deduced from:

\begin{equation}
  \label{eq:caseB}
  \frac{I(\text{Br}\beta)}{I(\text{Br}\alpha)}=\left[\frac{I(\text{Br}\beta)}{I(\text{Br}\alpha)}\right]^{*}10^{0.4A_{\rm V}\frac{N_{\rm H}}{A_{\rm V}}(C_{\rm ext}(\text{Br}\alpha)-C_{\rm ext}(\text{Br}\beta)}\ ,
\end{equation}

\noindent where $I(\text{Br}\beta)$/$I(\text{Br}\alpha)$ = 0.381$\pm$0.01966 is the attenuated line ratio measured from the NIRSpec spectrum of M17-B-PDR, $[I(\text{Br}\beta)/{I(\text{Br}\alpha)}]^{*}$ = 0.572 is the no-extinction value from \citet{1995MNRAS.272...41S}, $A_{\rm V}/N_{\rm H}$ = 5.3$\times$10$^{\rm -22}$~cm$^{\rm 2}$ is taken from \citet{2001ApJ...548..296W}, and the extinction cross sections $C_{\rm ext}(\text{Br}\alpha)$ = 2.816 and $C_{\rm ext}(\text{Br}\beta)$ = 5.694$\times$10$^{\rm -23}$~cm$^{\rm 2}$~H$^{\rm -1}$ when using the $R_{\rm V}$=5.5 curve from \citet{2001ApJ...548..296W}\footnote{$R_{\rm V}$=5.5 is chosen as it provides the most consistent results when performing spectral decomposition on the \textit{Spitzer} spectral cube of M17 \citep[][]{2018ApJ...858...67B} and is somewhat in line with \citet{1998A&A...329..161C}, who find $R_{\rm V}$=4.8.}. Furthermore, $[I(\text{Br}\beta)/{I(\text{Br}\alpha)}]^{*}$ is computed assuming an electron temperature of $T_{\rm e}$=10$^{\rm4}$~K and an electron density of $n_{\rm e}$=3$\times10^{\rm 3}$~cm$^{\rm -3}$. This exercise results in a considerable total visual extinction of $A_{\rm V}{\simeq}$8, which is agreeable with that found by \citet{2005ARep...49...36G} and \citet{2018ApJ...858...67B}.\\

\begin{figure}
    \centering
    \includegraphics[width=\linewidth]{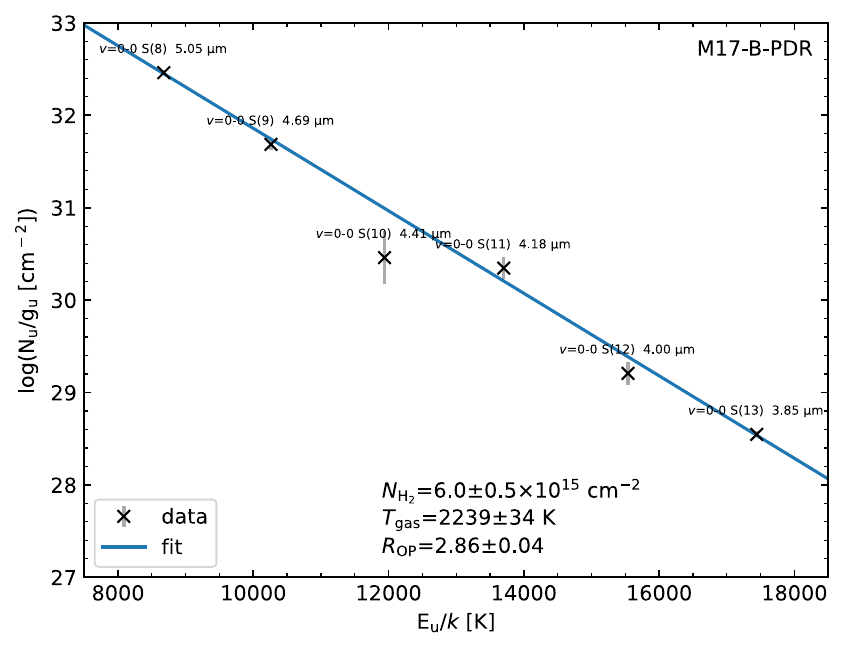}
    \caption{Pure rotational (not extinction-corrected) H$_{\rm 2}$ $v$=0-0 population diagram analysis of the spectrum of M17-B-PDR. The propagated statistical uncertainties have been indicated. The line (blue) shows the best fit, with the derived parameters given, including their uncertainties. See Section~\ref{subsubsec:other} for details and Section~\ref{subsec:lines} for a discussion.}
    \label{fig:population}
\end{figure}

\noindent\underline{H$_{\rm 2}$ lines}: Pure, $v$=0-0, rotational H$_{\rm 2}$ lines can be used to derive the molecular hydrogen column density and gas temperature from fits to population diagrams such as that presented in Figure~\ref{fig:population} \citep[e.g.,][]{2010ApJ...725..159F, 2018ApJ...858...67B}. A straight-line fit to the (not extinction-corrected) log-linear data in Figure~\ref{fig:population}, simultaneously optimizing the ortho-to-para ratio ($R_{\rm OP}$), gives a column density of $N_{\rm H_{\rm 2}}$=6.0$\pm$0.5$\times10^{\rm 15}$~cm$^{\rm -2}$, hot $T_{\rm H_{\rm 2}}$=2239$\pm$34~K gas, and an ortho-to-para ratio of $R_{\rm OP}=2.86\pm0.04$.

\section{Discussion}
\label{sec:discussion}

All targets display a spectrum with a rich and varying inventory of features that, taken together, help paint an astrophysical picture of the targeted areas. That is, each of the spectral components reveals something distinct about the astronomical environment. For example, the PAH bands provide direct insight into the level of excitation induced by the radiation field and of the chemical complexity that is present. Similarly, ice absorption bands give an impression of the molecular diversity encountered along the line-of-sight, and the atomic and molecular emission lines reveal gas densities and temperatures. Furthermore, the spectra probe distinct extended regions with different and changing morphologies spanning the aperture at unprecedented spatial resolution. Moreover, each target represents a different stage along the stellar life cycle, from star formation to stellar death. This changing astrophysical environment is reflected by the spectra through the appearance and disappearance of features, variations in relative intensities; the shape of the continuum; and band profiles.

What follows is a discussion of the PAH-related bands and other spectral features. The focus will be predominantly on the former as studying PAH and PAH-related emission is the main objective of JWST GO Cycle 1 program 1591. Density functional theory (DFT) computed PAH spectra that take \emph{anharmonicity} into account are relied upon to provide a backbone for the interpretation. Much of the discussion will be centered on the results obtained for M17-B-PDR in Section~\ref{subsec:spectra}.

\subsection{PAH Emision}
\label{subsubsec:PAHs}

The familiar PAH emission bands between 3-20~$\mu$m originate from PAH fundamental transitions, such as C-H and C-C stretches and bends, emitting from the first vibrational level, i.e., $v$=1-0. While the only PAH fundamentals that fall in the 1-5~$\mu$m range are the strong aromatic and aliphatic C-H stretching bands that dominate the 3.2-3.5~$\mu$m region, other, weaker, PAH emission features fill the entire 1-5~$\mu$m range producing the PAH-continuum \citep[e.g.,][]{1989ApJS...71..733A, 2021ApJ...917L..35A, 2023MolPhy2252936E}.

These weaker features are attributed to the effects of \emph{anharmonicity}, where vibrationally excited PAHs relax from higher vibrational levels by emitting \textit{1.} directly to the ground state, \textit{2.} step-wise down the vibrational ladder, and \textit{3.} in combinations with other vibrational modes (i.e., $\nu_\text{C-C}$+$\nu_\text{C-H}$). In addition, signatures of deuterated- and PAH-nitriles also fall in this wavelength range \citep[e.g.,][]{1989ApJS...71..733A, 2021ApJ...917L..35A}. 

DFT computations are utilized to aid in the interpretation of the PAH emission in the NIRSpec data. Previously, PAH spectra have been predominantly computed using the \emph{harmonic} approximation for frequencies and the double \emph{harmonic} approximation for intensities. This approach has shortcomings in predicting band positions and strengths for higher vibrational states that fall in, and dominate, the 1-5~$\mu$m region. With \textit{JWST} now providing unprecedented high-fidelity coverage of this wavelength range, commensurate theoretical methods that intrinsically include some level of \emph{anharmonicity} are required. Until recently, those methods could not be applied to molecules as large as PAHs. That has changed thanks to the development of novel techniques and the increase in computational power \citep[][]{2015JChPh.143v4314M, 2016JChPh.145h4313M, 2018JChPh.149m4302M, 2019A&A...632A..71C, 2022JPCA..126.3198M}. To that end, \citet{2023MolPhy2252936E} carried out DFT computations utilizing the Gaussian~16 quantum chemistry package \citep[][]{Gaussian16} with the B3LYP method in conjunction with the N07D \citep[][]{2008CPL...454..139B, 2010PCCP...12.1092B} basis set in order to obtain 1-30~$\mu$m \emph{anharmonic} absorption spectra for phenanthrene, pyrene, and pentacene. Also, new computations were performed that include unique singly-substituted, deuterated isotopologues as well as singly-substituted cyano (nitrile) side groups (Esposito~et~al. in prep.).

\subsubsection{The 3.2-3.6~\texorpdfstring{$\mu$}{u}m CH Stretching Region}
\label{subsubsec:pah33}

The 3.2-3.6~$\mu$m region is dominated by the strong aromatic CH stretching fundamental band at 3.3~$\mu$m and the moderately intense complex of overlapping bands between $\sim$3.38-3.6~$\mu$m that are associated with aliphatic CH stretching modes and combination band and overtone transitions from longer wavelength PAH modes \citep[e.g.,][]{1987ApJ...315L..61B, 1995A&A...299..835J, 2002A&A...388..639P,  2015ApJ...814...23M, 2022JPCA..126.3198M}.

The prominent 3.3~$\mu$m band originates from the aromatic C-H stretch in, relatively, smaller, neutral PAHs \citep[][]{1999ApJ...511L.115A}. Its profile and width have been characterized by \citet{2004ApJ...611..928V} into classes A, B1 and B2. Here, the bands peaking at 3.290, 3.293 and 3.297~$\mu$m are associated with class A, B1 and B2, respectively. For most astronomical PAH sources, the band is symmetric and of class A with a FWHM of $\sim$0.04~$\mu$m. Class B sources are less common and have a FWHM of $\sim$0.037~$\mu$m. In a number of cases, as also seen in some of the NIRSpec spectra presented here, the band is crowned with the Pf$\delta$ emission line at 3.296~$\mu$m. Emission around 3.3~$\mu$m is highly sensitive to anharmonicity and Fermi resonances \citep[][]{2015JChPh.143v4314M, 2018JChPh.149m4302M, 2022JPCA..126.3198M}, which can express itself as substructure in the band profile, e.g., its blue shoulder, satellite features at $\sim$3.4-3.6~$\mu$m and/or contribute to the plateau-like emission underlying the discrete band.

The distinct feature at 3.4~$\mu$m has been, besides being in part due to PAH anharmonicity, attributed to aliphatics, e.g., supra-hydrogenated PAHs. Post-AGB sources with extremely large 3.3/3.4~$\mu$m band ratios have an accompanying 6.9~$\mu$m band attributed to the methylene scissoring mode \citep[e.g.,][]{2017ApJ...850..165M}. Furthermore, the smooth evolution of the peak position of the PAH bands as a function of the effective temperature of the radiation field has been interpreted as a transition to a more aromatic nature of the emitting materials, i.e., loss of aliphatic side-groups \citep[][]{2005ApJ...621..831B, 2007ApJ...664.1144S}. Here, the NIRSpec data show a more typical astronomical 3.4~$\mu$m band, suggesting these sources are well in the aromatic domain. This is also consistent with extremely large 3.4/3.3~$\mu$m band ratios and a smooth evolution of peak position with effective temperature mainly being observed for class C sources, which are mostly evolved stars \citep[e.g.,][]{1992ApJ...387L..89G, 2007ApJ...662..389G, 2007ApJ...664.1144S}.

\subsubsection{The 1.6-1.7~\texorpdfstring{$\mu$}{u}m CH Stretch Overtone Region}
\label{subsubsec:overtone}

A number of astronomical studies have searched for the $v$=2-0 overtone transition of the 3.3~$\mu$m band, which, on first principles, is expected to fall near $\sim$1.67~$\mu$m \citep[][]{1992A&A...263..281M, 1993A&A...279L..45S, 1994ApJ...434L..15G}. A tentative detection was reported by \citet{1994ApJ...434L..15G} in IRAS21282+5050. Though as Figure~\ref{fig:overtone} reveals, any detection in the NIRSpec spectra can be heavily confused with strong emission lines. 

\citet[][]{2019A&A...632A..71C} showed, based on anharmonic DFT computations for eight PAHs in the C$_{\rm 14}$-C$_{\rm 20}$ size range, that PAH combination bands fall between 1.63-1.67~$\mu$m and pile up between 1.65-1.66~$\mu$m. However, overtones of the aromatic CH stretches are weak and rare. This is consistent with the results for phenanthrene, pyrene, and pentacene found here, where the first overtones of the aromatic CH stretches fall between 1.62-1.66~$\mu$m (6173-6024~cm$^{\rm -1}$), while significantly stronger combination bands with one CH stretch quanta and one other high-intensity mode quanta are spread across the 1.6-1.7~$\mu$m (6260-5882~cm$^{\rm -1}$) range. The CN and CD first overtones, however, are predicted at longer wavelengths.

Thus, on top of strong line confusion, any signal that can be isolated from the NIRSpec data and associated with PAHs cannot be directly attributed to solely the $v$=2-0 overtone of the 3.3~$\mu$m band.

\subsubsection{The 4.3-4.8~\texorpdfstring{$\mu$}{u}m Deuterated- and PAH-Nitriles Region}
\label{subsubsec:d+cn}

\paragraph{Deuterated PAHs}A wide range in the D/H gas ratio has been observed across the Galaxy \citep[e.g.,][]{2023ApJ...946...34F}. One possible explanation for this variation is the depletion of D onto PAHs via H exchange reactions in UV-rich environments \citep[e.g.,][]{2004oee..symp.....M}. As D is more strongly bound to a PAH than H, with the passage of time D enrichment builds, potentially leading to substantial PAH D/H ratios \citep[e.g.,][]{2004ApJ...604..252P, 2014ApJ...780..114O, 2016A&A...586A..65D, 2022ApJ...941..190O}.

Previous laboratory and harmonic DFT computational studies undertaken to determine the band positions of the C-D stretches showed that the aromatic stretches fall near 4.40~$\mu$m and the aliphatic C-D stretches near 4.65~$\mu$m. However, these positions shift slightly depending on PAH edge structure, extent of deuteration, and charge \citep[][]{1997JPCA..101.2414B, 2004ApJ...614..770H}. Many have searched for these signatures in astronomical data, with some reporting success \citep[e.g.,][]{2004ApJ...614..770H, 2004ApJ...604..252P, 2016A&A...586A..65D, 2020ApJS..251...12Y, 2022ApJ...941..190O}. Though, the predicted band positions of deuterated PAH coincide with strong emission lines from atoms and molecules and, before \textit{JWST}, observations were severely hampered by poor signal-to-noise and/or limited spectral resolution.
 
New anharmonic DFT computations to help identify signatures from deuterated PAHs in the NIRSpec data consider all the unique singly-substituted, deuterated isotopologues for phenanthrene, pyrene, and pentacene (Esposito~et al. in prep.). For phenanthrene and pyrene the aromatic CD stretch fundamental transitions are predicted to fall between 4.36-4.43~$\mu$m (2290-2130~cm$^{\rm -1}$), in line with \citep[][]{2004ApJ...614..770H}. However, for pentacene the aromatic C-D stretches fall between 4.68-4.69~$\mu$m.

\paragraph{PAH-nitriles}The recent discoveries of radio emission from benzonitrile (C$_{\rm 6}$H$_{\rm 5}$-C$\equiv$N), two cyanonaphthalenes (1-C$_{\rm 10}$H$_{\rm 7}$-C$\equiv$N and 2-C$_{\rm 10}$H$_{\rm 7}$-C$\equiv$N), and other small aromatic species containing C$\equiv$N in TMC-1 and, recently, other cold clouds \citep[][]{2018Sci...359..202M, 2021Sci...371.1265M, 2023arXiv230815951A} suggest cyano-PAHs (nitriles) could be an important subset of the astronomical PAH family.

\citet{2021ApJ...917L..35A} reported the C$\equiv$N stretch fundamental transitions of benzonitrile and other smaller, non-aromatic nitriles as falling between 4.46-4.5~$\mu$m. New \emph{anharmonic} computations of the C$\equiv$N stretch fundamental transitions for all unique isomers of singly-substituted cyano-benz[a]anthracene have a different range of 4.34-4.39~$\mu$m (2300-2275~cm$^{\rm -1}$; Esposito et.~al. in prep.). Consequently, this is also narrower than, and shifted from, the wavelength range for many deuterated PAHs (4.36-4.43~$\mu$m). For completeness, new harmonic DFT computations of the C$\equiv$N stretch for all unique isomers of singly-substituted large PAHs containing between 35-45 carbons atoms are consistent with the 4.34-4.39~$\mu$m range. Figure~\ref{fig:d+cn} indicates there is no clear-cut feature attributable to CN in this range.

Thus, given the limited range for the position of the computed C$\equiv$N stretch fundamental, any new feature observed between 4.36-4.69~$\mu$m that lies outside 4.34-4.39~$\mu$m may be due to the CD stretch fundamental transition.

Summing up, inspection of Figure~\ref{fig:d+cn} shows that despite huge improvements in spectral capabilities, the NIRSpec data do not allow for a straightforward assignment of any feature to deuterated- and/or PAH-nitriles. A decomposition method has to be employed to deal with the still present line confusion. For M17-B-PDR, this resulted in a potential feature at 4.65~$\mu$m that could be attributable to the aliphatic C-D stretch in deuterated PAHs, but, in agreement with \citet{2004ApJ...604..252P}, the absence of any reliable detection of a feature around 4.40~$\mu$m in M17 that could be attributed to the aromatic CD stretch is curious.

The [D/H] ratio determined for M17-B-PDR from the NIRSpec data, based on the aliphatic content alone, matches that reported by \citet{2004ApJ...604..252P} (0.31$\pm$0.127 vs. 0.36$\pm$0.08). However, this ratio is expected to depend on the exact location of the nebula probed and on the employed (emission) model \citep[][]{2014ApJ...780..114O, 2016A&A...586A..65D}.

\subsubsection{The 1-5~\texorpdfstring{$\mu$}{u}m PAH-Continuum}
\label{subsubsec:quasi}

Upon absorption of a UV photon, all PAH fundamental modes are temporarily vibrationally excited to some extent as the excitation energy is distributed and redistributed across the molecule during the relaxation process. The main relaxation channel is through the step-wise emission of infrared photons at frequencies corresponding to their fundamental vibrational frequencies as well as resonant combinations and overtones of those frequencies. The emission from this very large number of PAH transitions produces a continuum that spans the near-, mid-, and most of the far-infrared \citep[e.g.,][ and references therein]{1989ApJS...71..733A, 2021ApJ...917L..35A, 2023MolPhy2252936E}. The first astronomical reporting of such a continuum was by \citet[][]{1984ApJ...277..623S}.

PAHs containing some 40-70 carbon atoms, typical for the astronomical PAH population, have roughly 180-240 fundamental vibrational modes and thousands of combination band and overtone transitions. The density of these states for PAHs containing 40 carbon atoms is approximately 1$\times$10$^{\rm 23}$~states~cm$^{\rm -1}$ at 1~$\mu$m and 1$\times$10$^{\rm 7}$~states~cm$^{\rm -1}$ at 5~$\mu$m. For a 70 carbon atom sized PAH, these are 1$\times$10$^{\rm 31}$~states~cm$^{\rm -1}$ at 1~$\mu$m and 1$\times$10$^{\rm 10}$ states~cm$^{\rm -1}$ at 5~$\mu$m. \emph{For highly vibrationally excited molecules, symmetry determined selection rules are no longer as rigorous in emission as in the absorption case because the symmetry elements of the molecule in its ground state are no longer as well-defined and forbidden transitions (infrared dark states) become weakly allowed}. Thus, apart from expecting the intensity of each overtone band to drop as it progresses to higher order overtones, due to a decrease in intrinsic intensity of the participating vibrational transitions, it is impossible to predict, a-priori, their relative intensities \citep[][]{1945msms.book.....H}.

The 1-5~$\mu$m part of the PAH-continuum is largely ascribed to first-order combination bands and overtones of the strong PAH fundamental bands in the 5-15~$\mu$m region. Likewise, while dominated by the familiar PAH spectral signatures, the 5-15~$\mu$m region itself includes PAH-continuum emission from similar combination bands of the many longer wavelength, low frequency, PAH fundamentals described in \citet{2010ApJ...709...42R} and \citet{2011ApJ...729...64B}. Although significantly weaker than the strong fundamentals, blended emission from these contribute to the 5-15~$\mu$m region. This process continues into the far-infrared, with emission from combination bands of even longer wavelength bands continuously contributing to the emission at shorter wavelengths until the density of states drops such that continuum emission breaks up and falls away around 100~$\mu$m \citep[][]{2010ApJ...709...42R, 2011ApJ...729...64B}.

Although visible in earlier observations \citep[e.g.,][]{2001ApJ...551..807D, 2016AJ....151...93O}, NIRSpec now allows for the full characterization of the 1-5 $\mu$m continuum component of the astronomical PAH emission spectrum, revealing three broad bumps centered near 3.8, 4.04, and 4.34~$\mu$m whose positions suggest they could arise from combinations and overtones of the familiar 7.7, 7.8, and 8.6~$\mu$m PAH bands.

\subsection{Ices}
\label{subsec:ices}

The absorption by cold CO$_{\rm 2}$ water-ice in both M17 spectra and that of NGC1333-SVS3, with likely other ice bands as well (e.g., H$_{\rm 2}$O), will be spatially separated from the carriers of the other spectral components. As the low temperature of the CO$_{\rm 2}$-ice in M17-B-PDR suggests, the ices are likely associated with cold foreground molecular cloud material in the line-of-sight. In this specific case, it appears the light from the bright background PDR is making it partially through the foreground molecular material, in places being attenuated by denser filaments and cores.

The ices provide sites for ongoing chemistry, i.e., factories for creating more complex organic molecules like PAHs with alphatic side groups \citep[e.g.,][]{2011A&A...525A..93B, 2015ApJ...799...14C}. When eventually released as the PDR eats further and further into the molecular cloud material, this could help maintain any aliphatic fraction to support some 3.4~$\mu$m emission.

Ices will (partially) obscure any emission feature with which there is a spectral overlap. Thus, special care needs to be taken in the analysis and interpretation of such features. This includes features in the NIRSpec range, but also beyond, at mid-infrared wavelengths, including the major PAH bands. One specific example is CO-ice absorption at 4.67~$\mu$m obscuring potential emission from D- and/or PAH-nitriles at 4.65~$\mu$m \citep[][]{2022ApJ...941..190O}.

\subsection{Atomic and Molecular Lines}
\label{subsec:lines}

The rich inventory of atomic and molecular lines found in the NIRSpec spectra offer many diagnostic capabilities, e.g., photon- or shock-heating, and provide probes for the physical conditions of the gas, including its kinematics, cooling and chemistry \citep[][]{2004ARA&A..42..119V, 2022ARA&A..60..247W}.

The warm CO gas detected in IRAS21282+5050-RIM-BKSUB is likely associated with the PDR and consistent with detection at other wavelengths here and in other evolved stars \citep[][]{1993ApJ...411..266M, 2003MNRAS.344..262D}. The resolved fundamental P-R branch emission from CO would partially overlap with any potential deuterated- and/or cyano-PAH bands at 4.65~$\mu$m. Thus, separating such emission from the CO ro-vibrational structure, and other confused lines, requires care.

In the case of molecular hydrogen, collisions keep the lowest rotational levels of ($v$=0, $J{<}$5) in thermal equilibrium and thus provide a good indicator of the gas temperature and insight into the ortho-to-para ratio \citep[e.g.,][]{1989ESASP.290..269S, 2005SSRv..119...71H}. However, the population diagram in Figure~\ref{fig:population} probes the $J>5$ S(8) to S(13) lines. Thus, the levels in the NIRSpec range are excited through UV pumping \citep[][]{2005SSRv..119...71H}, which is further confirmed by the 1-0 S(1)/2-1 S(1) ratio of 2.0 \citep[see e.g.,][]{2002MNRAS.333..721B}. Furthermore, in equilibrium the ortho-to-para-ratio $R_{\rm OP}$ is expected to be 3, with 2.86 found here for M17-B-PDR. Therefore, the high temperature for for M17-B-PDR needs to be interpreted as an excitation temperature.

\subsection{Variations along the Stellar Life Cycle}
\label{subsec:lifecycle}

Along the stellar life cycle gas and dust flow through ever-changing environments of varying densities, temperatures and radiation fields. Some environments will promote an increase in chemical complexity, be it through gas-phase reactions or grain surface chemistry, while others will reduce chemical complexity and only the most robust chemical species and particles survive.

Besides reflecting the current astrophysical environment, infrared spectra also capture the history of the gas and dust. The NIRSpec spectra presented here sample a number of key stages along the low-mass stellar life cycle. Although similar overall, there are a number of striking and subtle differences. With the emphasis on PAH and related features, PAHs are not only silent witnesses of local conditions, they also help shape their environment. It is this interplay that simultaneously reveals characteristics of the PAH population itself and the astrophysical environment in which they reside \citep[e.g,,][]{2008ARA&A..46..289T}.

\section{Summary and Conclusions}
\label{sec:conclusions}

The James Webb Space Telescope (\textit{JWST}) represents a giant leap in infrared spectral-spatial fidelity and sensitivity. A first look at the  1-5~$\mu$m observations from \textit{JWST} GO Cycle~1 program 1591 has been presented here. Using NIRSpec-IFU, the program targets 7 objects along the low-mass stellar life cycle with PAH emission. 

Aperture positions were chosen such that they capture a varying morphology with gradients, ridges and distinct features, which allows for a spatial analysis across single extended objects. Some of the spectra show notable variation within their extraction aperture. These variations are consistent with the spatial morphology that can be observed in their 3.3~$\mu$m PAH maps and line up well with what is seen in color-composite imagery that reveal the overall, larger scale, structure of the targeted regions.

This first look paper explores the spectrum of each target extracted from a 1.5$^{\prime\prime}$ radius sized circular aperture centered on the field-of-view. All spectra show a wealth of features, including the 3.3 and 3.4~$\mu$m PAH complex, the PAH-continuum, and a vast array of atomic and molecular emission lines including HI, H$_{\rm 2}$, and likely an array of other species. Some spectra also show absorption by CO$_{\rm 2}$- and H$_{\rm 2}$O-ice and emission from warm gas-phase CO. 

For most sources, the prominent PAH CH stretch band is symmetric and of Class A, peaking at 3.29~$\mu$m (FWHM=0.04~$\mu$m). Its anticipated $v$=2-0 overtone at 1.68~$\mu$m is confused by line emission in all targets. The 3.4~$\mu$m band shows two varying components that can be ascribed to aliphatic CH stretches and/or hot bands of the aromatic CH stretch.

The PAH-continuum, characterized for the first time, spans most of the 1-5~$\mu$m region. Rising slowly from $\sim$1~$\mu$m, it jumps near 3.2~$\mu$m and slowly continues to rise out to 5~$\mu$m. Broad bumps at roughly 3.8, 4.04, and 4.34~$\mu$m add structure for all but the spectra of NGC7027-EDGE and BD+303639-RIM, where they could be hidden by the forest of emission lines.

Despite \textit{JWST}'s spectral resolution and sensitivity, identifying possible emission from deuterated- and/or cyano-PAHs remains challenging. For the bright-PDR position in M17, a multi-component, decomposition was carried out. While this indeed suggests that an aliphatic CD stretch feature centered at 4.65~$\mu$m (FWHM=0.02~$\mu$m) is present, consistent with earlier work, there is no hint for the aromatic CD stretch at 4.40~$\mu$m. The latter is also true for any cyano-PAH feature between 4.34-4.39~$\mu$m. Thus, despite huge improvements in spectral resolution and sensitivity, at this stage of analysis, these NIRSpec spectra do not allow for a straightforward assignment of any feature due to deuterated- and/or cyano-PAHs.

The CO$_{\rm 2}$ absorption band seen in the M17-B-PDR spectrum is well matched with 10:1 H$_{\rm 2}$O:CO$_{\rm 2}$ ice at 10~K. A two component fit to its $v$=0 pure rotational molecular hydrogen population diagram shows high-temperature gas that, given the high-J rotational lines, is originating from UV-pumped gas. M17-B-PDR also shows a large number of hydrogen recombination lines, with the atomic hydrogen Pfund series running from level 10 to well into the 30-ies. Considering Br$\beta$/Br$\alpha$=0.382$\pm$0.2002 and Case B recombination, an estimated total visual extinction of $A_{\rm V}{\eqsim}$8 is found for M17-B-PDR.

Lastly, the ro-vibrational structure of CO seen in the spectrum of IRAS21282+5050 is well-matched with warm, 258~K, gas.\\

\section{Software and third party data repository citations}
\label{sec:cite}

The JWST data presented in this paper were obtained from the Mikulski Archive for Space Telescopes (MAST) at the Space Telescope Science Institute (STScI). The STScI-reduced observations can be accessed via \dataset[https://doi.org/10.17909/x16g-g718]{https://doi.org/10.17909/x16g-g718}. STScI is operated by the Association of Universities for Research in Astronomy, Inc., under NASA contract NAS5–26555. Support to MAST for these data is provided by the NASA Office of Space Science via grant NAG5–7584 and by other grants and contracts.

Where indicated, color composite images were created from observations made with the NASA/ESA Hubble Space Telescope. Those data were obtained from the Hubble Legacy Archive, which is a collaboration between the Space Telescope Science Institute (STScI/NASA), the Space Telescope European Coordinating Facility (ST-ECF/ESA) and the Canadian Astronomy Data Centre (CADC/NRC/CSA). 

Density functional theory computations were performed using the Gaussian~16 software suite \citep[][]{Gaussian16}.

\begin{acknowledgments}

This work would not have been possible without T.J.L., who passed away during the development of this project. His guiding hand and expertise in computational chemistry as applied to astronomy and astrophysics will be deeply missed by this team and all in the community.\\

This project is based on observations made with \textit{JWST}. Financial support from the Space Science Telescope Institute is acknowledged (GO-01591). C.B. is grateful for an appointment at NASA Ames Research Center through the San Jos\'e State University Research Foundation (80NSSC22M0107). C.B., J.D.B., L.J.A. and A.M. acknowledge support from the Internal Scientist Funding Model (ISFM) Laboratory Astrophysics Directed Work Package at NASA Ames (22-A22ISFM-0009). L.J.A., J.D.B. and A.M. are thankful for an appointment at NASA Ames Research Center through the Bay Area Environmental Research Institute (80NSSC19M0193). V.J.E.'s research has been supported by an appointment to the NASA Postdoctoral Program at NASA Ames Research Center, administered by the Oak Ridge Associated Universities through a contract with NASA. R.C.F. is grateful for support from NASA grants NNX17AH15G and NNH22ZHA004C, the University of Mississippi's College of Liberal Arts, and the Mississippi Center for Supercomputing Research, which is funded in part by NSF Grant OIA-1757220. E.P. acknowledges support from the University of Western Ontario, the Institute for Earth and Space Exploration, the Canadian Space Agency, and the Natural Sciences and Engineering Research Council of Canada. Special thanks go to program coordinator W.J.~Skipper, NIRSpec reviewer K.~Namisha and the \textit{JWST} Help Desk, notably T.~Keyes, for their technical support. Lastly, the anonymous referee is gratefully acknowledged for their valuable insight and feedback.

\end{acknowledgments}

\vspace{5mm}
\facilities{JWST(NIRSpec)}

\software{Astropy \citep[][]{2013A&A...558A..33A, 2018AJ....156..123A}, JWST Calibration Pipeline \citep[][]{bushouse_howard_2022_7182148}}

\bibliography{aamnem99,bibliography}{}
\bibliographystyle{aasjournal}

\end{document}